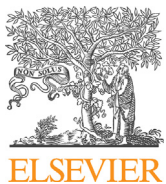
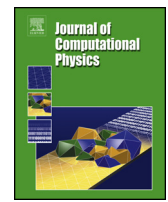

# Applying Bayesian optimization with Gaussian process regression to computational fluid dynamics problems

Y. Morita [a,c,∗], S. Rezaeiravesh [a,b,∗], N. Tabatabaei [a,b], R. Vinuesa [a,b], K. Fukagata [c], P. Schlatter [a,b]

[a] *SimEx/FLOW, Engineering Mechanics, KTH Royal Institute of Technology, Stockholm, Sweden*
[b] *Swedish e-Science Research Centre (SeRC), Stockholm, Sweden*
[c] *Department of Mechanical Engineering, Keio University, Yokohama, Japan*

## A R T I C L E   I N F O

## A B S T R A C T

Bayesian optimization (BO) based on Gaussian process regression (GPR) is applied to different CFD (computational fluid dynamics) problems which can be of practical relevance. The problems are i) shape optimization in a lid-driven cavity to minimize or maximize the energy dissipation, ii) shape optimization of the wall of a channel flow in order to obtain a desired pressure-gradient distribution along the edge of the turbulent boundary layer formed on the other wall, and finally, iii) optimization of the controlling parameters of a spoiler-ice model to attain the aerodynamic characteristics of the airfoil with an actual surface ice. The diversity of the optimization problems, independence of the optimization approach from any adjoint information, the ease of employing different CFD solvers in the optimization loop, and more importantly, the relatively small number of the required flow simulations reveal the flexibility, efficiency, and versatility of the BO-GPR approach in CFD applications. It is shown that to ensure finding the global optimum of the design parameters of the size up to 8, less than 90 executions of the CFD solvers are needed. Furthermore, it is observed that the number of flow simulations does not significantly increase with the number of design parameters. The associated computational cost of these simulations can be affordable for many optimization cases with practical relevance.

© 2021 The Author(s). Published by Elsevier Inc. This is an open access article under the CC BY license (http://creativecommons.org/licenses/by/4.0/).

## 1. Introduction

In computational fluid dynamics (CFD), the Navier–Stokes equations or variations of them are numerically solved in order to simulate fluid flows. At most of the Reynolds numbers associated to environmental flows and relevant to engineering applications, flows are turbulent, see *e.g.* Ref. [14]. Various approaches have been developed to numerically simulate turbulent flows, ranging from low-fidelity Reynolds-averaged Navier–Stokes (RANS) simulation to high-fidelity approaches including large eddy simulation (LES) and direct numerical simulation (DNS), see Ref. [63]. Moving from RANS toward DNS, the influence of modeling is reduced, the role of numerics becomes more dominant, and above all, the computational cost increases [68]. In many applications where the fluid flows are involved, we need to optimize different parameters in order to meet (within a margin) a certain set of objectives while satisfying a set of constraints, see Refs. [76,17] and the references therein. The aim of the present paper is to illustrate how Bayesian optimization (BO) [67,21], which has been mostly utilized in the field of machine learning and data sciences, can be applied to different types of optimization problems arising in CFD. Previous application of Bayesian optimization to CFD problems is limited to a few studies, see Refs. [73,44,38]; therefore, the diversity of the applications considered in the present study can reveal the flexibility of the method as well as shed light on its pros and cons.

---

* Principal corresponding author at: SimEx/FLOW, Engineering Mechanics, KTH Royal Institute of Technology, Stockholm, Sweden.
*E-mail addresses:* morita@kth.se (Y. Morita), salehr@kth.se (S. Rezaeiravesh), nargest@mech.kth.se (N. Tabatabaei), rvinuesa@mech.kth.se (R. Vinuesa), fukagata@mech.keio.ac.jp (K. Fukagata), pschlatt@mech.kth.se (P. Schlatter).

To put the work in context, we shortly review the main approaches which have been employed for the purpose of optimization in CFD. Such approaches are iterative and can, in general, be divided into gradient-based, gradient-free, and gradient-enhanced, see Refs. [46,76]. The gradient-based methods are the appropriate choice when the evaluation of the derivatives of the cost function with respect to the design parameters is computationally efficient. At each iteration, the first-order gradient is needed to determine the searching direction and the second-order gradients (*i.e.* Hessian matrix) or the approximations of them are required to obtain the optimal step size, see *e.g.* Ref. [46]. Therefore, depending on the approach, a minimum order of smoothness for the objective function is necessary. The gradients can be computed by automatic-differentiation (finite differences), or more efficiently by solving forward or adjoint sensitivity equations. The use of the latter is recommended when the number of design parameters is more than the number of objectives, see [46,28,76]. As its main advantage, the gradient-based optimizations exhibit fast convergence and can handle large numbers of design parameters. However, it is always likely to be trapped by local optima. Besides this, in the cases of unsteady Navier–Stokes equations the need for saving the forward solution fields which are required to solve the adjoint equations may lead to memory issues. To rectify these issues, extra treatments are needed, see [3,29]. Various types of application of gradient-based methods for optimization in CFD can be found in the literature, including shape optimization [41], optimization of initial conditions for fast transition to turbulence [32], and topology optimization in turbulent flows using the RANS approach [17].

In gradient-free approaches, the simulator (here, the CFD solver) is treated as a blackbox and run to evaluate realizations of the response at the samples of the design parameters. This approach which goes back to Box and Draper [7] can be implemented using different methods, see Refs. [33,20]. The simplest method is the grid search, where a set of manually selected samples are tested to find the optimum. Evolutionary algorithms [2], which mimic the nature's survival by iteratively simulating "selection and mutation" of the design parameters, are also considered as an effective approach; see *e.g.* Ref. [82] where the algorithm is applied to modify the RANS stress–strain relationship. However, the most frequently-used method of this type in CFD is called the response-surface method (RSM), in which a metamodel or surrogate is constructed using a finite set of training data comprised of parameter samples and corresponding responses. The size of the training data set is determined as a balance between the required accuracy of the surrogate and the cost involved in running the simulator. In many CFD applications, Gaussian process regression (GPR), or the Kriging method [57,27], has been employed to construct the surrogate. As detailed in Section 2, GPR is among the non-parametric surrogates, naturally allows for noisy (uncertain) data, and more importantly, can estimate the uncertainties involved in their predictions. Examples of RSM-based optimization can be found, for instance, in Viana et al. [78]. Contrary to the gradient-based methods, RSM is suitable for finding the global optima, but it suffers from the curse of dimensionality. To reduce the required number of samples for high-dimensional parameters yet constructing an accurate response surface, gradient information can be added to the training data, see the review by Laurent et al. [35]. In particular, the gradient-enhanced Kriging (GEK) has been extensively used in the last two decades for optimization in CFD, see for instance, [11,34]. The study by Laurenceau et al. [34] showed that at large number of parameters (=45), GEK outperforms RSM-GPR, while at small number of parameters (=6), inclusion of the gradients makes no significant improvement in the computational performance of the optimization.

In the present study, the Bayesian optimization based on Gaussian process regression, hereafter referred to as BO-GPR, is employed, see Refs. [21,67]. This approach can be classified among the gradient-free approaches, although a gradient-enhanced version of it has also been proposed, see Ref. [84]. In BO, a sequence of samples for the design parameters is taken which converges to the global optimum. Therefore, as a main difference with the RSM approach, the surrogate in BO is sequentially updated. That is, in fact, the clear connection of BO to the Bayesian formalism, see *e.g.* Ref. [69]. As detailed in Section 2, at each iteration, the decision about the next sample is taken based on combining the exploitation and exploration, where the latter directly takes into account the predictive uncertainty of the surrogate. This active involvement of the uncertainties in the algorithm is another distinguishing characteristic of the BO approach over RSM methods. In general, the predictive uncertainties are a combination of the uncertain CFD data and the constructed surrogate. The use of the BO is a timely choice considering the application of the uncertainty quantification (UQ) [69] in different aspects of CFD and numerical simulation of turbulence over the last decade, see *e.g.* Refs. [4,48,85,58,59,61]. In fact, most of those studies are classified as UQ forward problems, the outcomes of which can be nicely employed in BO, which is, in fact, an inverse UQ problem. Having the design parameters defined over an admissible space, the BO can be non-intrusively linked to any CFD solver. This provides a great flexibility noting that a CFD solver is not necessarily equipped with adjoint solvers to compute gradients, and in any case, computing the adjoint sensitivities could be expensive.

The application of BO in CFD has been very limited. Talnikar et al. [73] developed a parallel version of BO to minimize the drag in a turbulent channel flow simulated by LES. Nabae et al. [44] used BO to optimize the phase-speed of streamwise traveling wave-like wall deformation for drag reduction in a turbulent channel flow simulated by DNS. More recently, BO was employed by Mahfoze et al. [38] to optimize the blowing amplitude and coverage in order to reduce the friction drag to achieve a certain net-power saving in a zero-pressure-gradient turbulent boundary layer (TBL) simulated by DNS. To study different aspects of BO, a diverse set of optimization problems are considered in the present paper. The objectives, number of design parameters, and CFD solver are different in these problems.

This paper is structured as follows. In Section 2, a detailed review is given to the theoretical aspects of the adopted Bayesian optimization methodology. In Section 3, BO is applied to optimize the initial conditions of a set of coupled non-linear ordinary differential equations so that an energy norm at some later time is maximized. As the second example, shape optimization of a laminar lid-driven cavity flow is studied in Section 4, where the objective is to either minimize or maximize the energy dissipation. Another shape optimization which has practical applications, is considered in Section 5. The objective is to optimize the shape of a wall boundary of a channel, so that the streamwise pressure gradient of the turbulent boundary layer (TBL) over the other wall matches a given profile. Section 6 is devoted to the optimization of a spoiler-ice airfoil profile. The parameters controlling the configuration of two spoilers which are located on the pressure and suction sides of the airfoil are optimized so that the distribution of the pressure on the airfoil surface becomes the same as if a real ice was formed on the front part of the airfoil. In the end, concluding remarks are provided in Section 7.

## 2. Bayesian optimization

Consider the set of design parameters $\mathbf{q} = \{q_1, q_2, \cdots, q_d\}$ which are allowed to vary over the admissible space $\mathbb{Q} \subset \mathbb{R}^d$ (note that $\mathbb{Q}$ is not the set of rational numbers in this paper). Running the simulator, *i.e.* the CFD code, realizations for the quantities of interest (QoIs) or responses are obtained, which in general can be noisy. The aim of the optimization is to minimize (or maximize) an objective function $r(\mathbf{q})$ which depends on the simulator outputs over space $\mathbb{Q}$, and find optimal parameter $\mathbf{q}_{\text{opt}}$,

$$\mathbf{q}_{\text{opt}} = \arg\min_{\mathbf{q} \in \mathbb{Q}} r(\mathbf{q}). \tag{1}$$

Ideally, a functional dependency between the observed values $r$ and parameters $\mathbf{q}$ could be established as [57,27],

$$r(\mathbf{q}) = f(\mathbf{q}) + \boldsymbol{\varepsilon}, \tag{2}$$

where, $\boldsymbol{\varepsilon}$ denotes the observation noise, and without loss of generality it is assumed to be identically distributed for all samples as $\mathcal{N}(0, \sigma^2)$. However, the complexities in the flow solver make acquiring a closed form for $f(\mathbf{q})$ infeasible. Alternatively, a surrogate $\tilde{f}(\mathbf{q})$ can be constructed adopting a non-intrusive approach in which the simulator is treated as a blackbox. To this end, a finite set of training data $\mathcal{D} = \{(\mathbf{Q}^{(i)}, \mathbf{R}^{(i)})\}_{i=1}^n$ is employed, where $\mathbf{Q}$ and $\mathbf{R}$ are respectively samples of $\mathbf{q}$ and corresponding realizations of $r$. In general, the number of the training data, $n$, is limited due to the cost of running the simulator.

The two main components of the Bayesian optimization, *i.e.* the choice of method for constructing $\tilde{f}(\mathbf{q})$, and drawing the sequence of training samples, are now shortly reviewed. The Gaussian process regression (GPR) [57,27] is employed to build $\tilde{f}(\mathbf{q})$. In this approach, $\tilde{f}(\mathbf{q})$ is assumed to be random for which a prior distribution in the form of a Gaussian process is assumed. A Gaussian process $\mathcal{GP}$ is a multi-variate Gaussian distribution defined as,

$$\tilde{f}(\mathbf{q}) \sim \mathcal{GP}\left(m(\mathbf{q}), c(\mathbf{q}, \mathbf{q}'; \boldsymbol{\Theta}_\varepsilon, \boldsymbol{\beta})\right), \tag{3}$$

in which the mean and covariance are respectively given by,

$$m(\mathbf{q}) = \mathbb{E}[\tilde{f}(\mathbf{q})], \tag{4}$$

$$c(\mathbf{q}, \mathbf{q}'; \boldsymbol{\Theta}_\varepsilon, \boldsymbol{\beta}) = \mathbb{E}[(\tilde{f}(\mathbf{q}) - m(\mathbf{q}))(\tilde{f}(\mathbf{q}') - m(\mathbf{q}'))]. \tag{5}$$

In these definitions, $\mathbb{E}[\cdot]$ denotes the expected value, $\boldsymbol{\Theta}_\varepsilon$ specify the parameters building the noise structure, and $\boldsymbol{\beta}$ are the hyperparameters in the kernel function which represents the covariance of $\tilde{f}(\mathbf{q})$. Given the training data $\mathcal{D}$ along with the associated observation uncertainty, the posterior and posterior predictive of $\tilde{f}(\mathbf{q})$ (conditioned on $\mathcal{D}$ and $\boldsymbol{\Theta}_\varepsilon$) can be inferred. Simultaneously, the hyperparameters $\boldsymbol{\beta}$ are optimized. The posterior-predictive of $\tilde{f}(\mathbf{q})$ at a set of test samples $\mathbf{Q}^* \in \mathbb{Q}$, where $\mathbf{Q}^* = \{\mathbf{q}^{*(i)}\}_{i=1}^{n^*}$, has a multivariate Gaussian distribution $\mathcal{N}(m(\mathbf{R}^*|\mathcal{D}, \boldsymbol{\Theta}_\varepsilon, \mathbf{Q}^*), v(\mathbf{R}^*|\mathcal{D}, \boldsymbol{\Theta}_\varepsilon, \mathbf{Q}^*))$, where,

$$m(\mathbf{R}^*|\mathcal{D}, \boldsymbol{\Theta}_\varepsilon, \mathbf{Q}^*) = \mathbf{C}(\mathbf{Q}^*, \mathbf{Q})(\mathbf{C}(\mathbf{Q}, \mathbf{Q}) + \mathbf{C}_N)^{-1}\mathbf{R}^T, \tag{6}$$

$$v(\mathbf{R}^*|\mathcal{D}, \boldsymbol{\Theta}_\varepsilon, \mathbf{Q}^*) = \mathbf{C}(\mathbf{Q}^*, \mathbf{Q}^*) - \mathbf{C}(\mathbf{Q}^*, \mathbf{Q})(\mathbf{C}(\mathbf{Q}, \mathbf{Q}) + \mathbf{C}_N)^{-1}\mathbf{C}(\mathbf{Q}, \mathbf{Q}^*). \tag{7}$$

Here, without loss of generality, the mean of $\tilde{f}(\mathbf{q})$ in Eq. (3) is assumed to be zero, $\mathbf{R} = [r^{(1)}, r^{(2)}, \cdots, r^{(n)}]$, $\mathbf{C}(\mathbf{Q}, \mathbf{Q}')$ is a $n \times n'$ matrix where $[\mathbf{C}(\mathbf{Q}, \mathbf{Q}')]_{ij} = c(\mathbf{q}^{(i)}, \mathbf{q}'^{(j)})$ and $c(\cdot, \cdot)$ is a kernel function dependent on hyperparameters $\boldsymbol{\beta}$. Moreover, $\mathbf{C}_N$ is the covariance matrix of the noise in the training data. In this derivation, the noise samples are assumed to be independent and identically distributed (iid) as $\varepsilon \sim \mathcal{N}(0, \sigma^2)$. A more general case in which the noise is allowed to be observation-dependent has been developed by Goldberg et al. [24], and recently applied for the purpose of uncertainty quantification in CFD by Rezaeiravesh et al. [59]. In practice, the observational uncertainties can, for instance, be due to the convergence limits imposed when solving discretized equations and also finite time-averaging in unsteady flows. Including these uncertainties in the BO-GPR requires considering two points. 1. The computed optimum is, in general, uncertain and it may not be possible to obtain a noise-free value. In practice, when maximizing the acquisition function, the expected value of the posterior predictive for the noisy response can be used, see Ref. [26]. 2. The standard acquisition functions, such as expected improvement, Eq. (8), may not provide adequate exploration over the parameter space, therefore, the global optima may not be found or the convergence rate is very slow [77]. Recently, Letham et al. [36] has proposed a modified version of the expected improvement function for handling the noisy data (there are other acquisition functions for noisy data which have been reviewed in Ref. [36]). Implementing such modified acquisition functions for BO-GPR in CFD will be considered in a future extension of the present work.

Now, a short overview is given on how the sequence of samples for $\mathbf{q}$ can be drawn from $\mathbb{Q}$ so that they converge to $\mathbf{q}_{\text{opt}}$. At the $k$-th iteration of optimization the training data set is $\mathcal{D}_{1:k} := \{(\mathbf{q}^{(i)}, r^{(i)})\}_{i=1}^{k}$. The decision about the next sample $\mathbf{q}^{(k+1)}$ is made through maximizing an acquisition function (ACF) $\alpha(\mathbf{q}; \mathcal{D}_{1:k})$, see *e.g.* Ref. [21]. The most popular ACF is the expected improvement (EI) which was first suggested by Močkus [40] and then utilized in BO by Jones et al. [31]. The EI for the minimization problem is defined as,

$$\text{EI}(\mathbf{q}) := \mathbb{E}[I(\mathbf{q})] = \mathbb{E}[\max\left(r(\mathbf{q}^{\dagger}) - r(\mathbf{q}), 0\right)], \tag{8}$$

where $\mathbf{q}^{\dagger} = \arg\min_{\mathbf{q} \in \mathcal{D}_{1:k}} r(\mathbf{q})$, denotes the best estimate for optimum among the $k$ observations (iterations). Note that the improvement $I(\mathbf{q})$ is random since $r(\mathbf{q})$ is so, see Eq. (2). In the cases of using a standard GPR for $\tilde{f}(\mathbf{q})$, Jones et al. [31] derived a closed-form expression for the EI which reads as,

$$\text{EI}(\mathbf{q}) = \begin{cases} \left(r(\mathbf{q}^{\dagger}) - m(\mathbf{q}) + \xi\right) \Phi(\zeta) + s(\mathbf{q})\varphi(\zeta), & s(\mathbf{q}) > 0 \\ 0, & s(\mathbf{q}) = 0 \end{cases}, \tag{9}$$

where $s(\mathbf{q}) = \sqrt{v(\mathbf{q})}$, and $m(\mathbf{q})$ and $v(\mathbf{q})$ are the mean and standard deviation of the posterior predictive of the GPR at any $\mathbf{q}$, as given by Eqs. (6) and (7), respectively. Further, $\Phi(\zeta)$ and $\varphi(\zeta)$ respectively represent the CDF (cumulative distribution function) and PDF (probability density function) of the standard Gaussian distribution for $\zeta$ defined by,

$$\zeta = \begin{cases} \left(r(\mathbf{q}^{\dagger}) - m(\mathbf{q}) + \xi\right)/s(\mathbf{q}), & s(\mathbf{q}) > 0 \\ 0, & s(\mathbf{q}) = 0 \end{cases}. \tag{10}$$

In Eq. (9), the first term in the summation specifies exploitation, the use of the best value so far, and the second term identifies exploration parts of the $\mathbb{Q}$ where the surrogate has highest uncertainty. The ad-hoc parameter $\xi$ can be used in practice as the controller of the exploitation in comparison with exploration.

In the present study, the Bayesian optimization algorithm based on GPR is implemented using the `Python` libraries `GPy` [25] and `GPyOpt` [75]. The developed `Python` codes are non-intrusively linked to the CFD solvers, Nek5000 [19] and OpenFOAM [83] through appropriate `bash` drivers. As a result, the whole optimization loop is fully automated. All optimizations started from a random sample for $\mathbf{q}$ taken from $\mathbb{Q}$. For constructing the kernel in the GPR surrogates, the Matern52 function is used, see *e.g.* [57,27]. Note that in general, the choice of the kernel function can affect the convergence of the BO-GPR [62]. Our tests showed that Matern52 kernel could lead to a faster convergence for a set of tested problems. To optimize the GPR hyperparameters, the Broyden-Fletcher-Goldfarb-Shanno (BFGS) algorithm is utilized, see *e.g.* [46]. In Appendix B, a short discussion is made on the computational time required for BO-GPR and CFD.

Recognizing the convergence of the BO and hence imposing a stopping criterion for that can be, in general, not so straightforward, see *e.g.* [9,66]. In fact, this is a common characteristic of all gradient-free optimizations. In some cases, monitoring $\|\mathbf{q}^{(k)} - \mathbf{q}^{(k-1)}\|_2$ at the $k$-th iteration can help. But noting that the exploration guides the samples to different parts of the admissible space $\mathbb{Q}$, obtaining a small difference between the successive samples, especially for large-dimensional parameters, is difficult to achieve in practice. We have observed that a more reliable measure is tracking the best value of the objective, *i.e.* $r(\mathbf{q}^{\dagger})$ where $\mathbf{q}^{\dagger} = \arg\min_{\mathbf{q} \in \mathcal{D}_{1:k}} r(\mathbf{q})$. If this measure remains unchanged after a sufficiently large $n$, then we can consider the optimization to be converged (this is called the identification step, according to Ref. [30]). However, the value of $n$ is not known *a priori* and it is basically limited by the computational cost of the simulations and available computational budget. Despite this, examining the response surface constructed by the surrogate can help identifying parts of $\mathbb{Q}$ with insufficient samples at which the predictive uncertainty of the surrogate is high. Guiding the sequence of samples toward those parts can be, for instance, done by adjusting $\xi$ in Eq. (9). Despite this, the cases studied in Sections 5 and 6 are not directly affected by the above discussion, since the optimum value for the objective is the deviation of a computed QoI from a target value, and is therefore *a priori* known (albeit potentially not exactly achievable given the design space). In Section 4, where this is not the case, monitoring the best value found when sufficiently many samples are drawn will help.

## 3. An introductory example

In many applications in fluid dynamics, in particular hydrodynamic stability, we are interested in finding the optimum initial condition which leads to fastest growth of the unstable modes and hence fastest transition from laminar state to turbulence. Kerswell et al. [32] have considered this problem through using adjoint-based optimization. As a "toy" problem to examine their approach, Kerswell et al. [32] introduced the following dynamical system comprised of two ordinary differential equations (ODE),

$$\frac{d\mathbf{x}}{dt} = \begin{bmatrix} -x_1 + 10x_2 \\ x_2(10\exp(-x_1^2/100) - x_2)(x_2 - 1)) \end{bmatrix}, \quad (11)$$

with the initial condition $\mathbf{x}(0) = \mathbf{x}_0$. The objective is to find the initial data $\mathbf{x}_0$ parameterized as $\mathbf{x}_0 = a(\cos\pi\theta, \sin\pi\theta)$ (with fixed $a$) which maximizes $\|\mathbf{x}(T)\|_2^2/a^2$, the normalized energy gain at time $T$. Therefore, the design parameter is $q = \theta$ which is allowed to vary over the admissible range $\mathbb{Q} = [0, 2)$. Based the above parameterization, all solutions $\mathbf{x}(t)$ start from the same distance $\|\mathbf{x}_0\|_2 = a$. Depending on the value of $a$, the trajectories for $\mathbf{x}(t)$ will end up at different attractors of the dynamical system at time $T$. In the gradient-based optimization adopted in [32], the adjoint equation of Eq. (11) is derived using the Lagrange multiplier approach and is then solved backward from $t = T$ to $t = 0$ in each iteration of the optimization. Here, we apply the BO algorithm explained in Section 2 to find the optimum $\theta$ for different values of $a$ adopted in [32]. For all values of $a$, the BO starts from the same initial samples $\theta = 0.1$ and 1.5. Our further study showed that the initial samples have no impact on the final optimal values, but they can slightly affect the sequence of samples and hence convergence of the BO. As shown in Fig. 1, the computed optima by the BO-GPR approach are accurate (compared to the true values) and the same as those in Ref. [32] where an adjoint method was employed. This is a significant advantage of the BO-GPR method, noting that solving the same problem with the adjoint-based method could be more involved and for some values of $a$ the convergence of the algorithm would be a bit difficult to reach due to numerical instabilities. On the other hand, the convergence in all cases shown in Fig. 1 is achieved with less than 15 samples. The history of the optimization is presented in Figs. A.15 and A.16 in Appendix A.

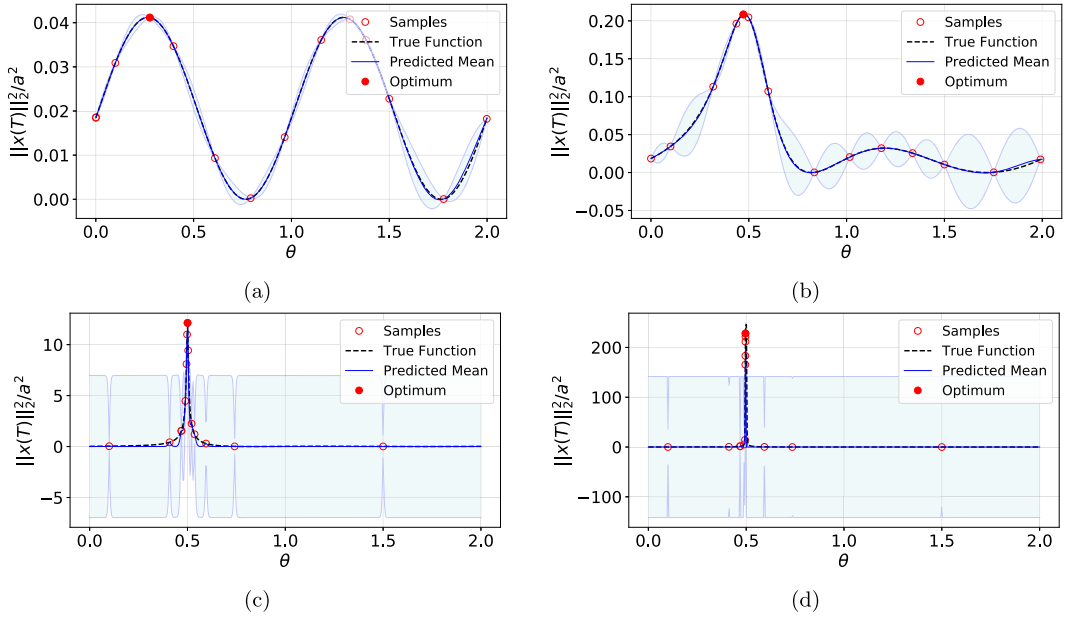

**Fig. 1.** Gain $\|\mathbf{x}(T)\|_2^2/a^2$ plotted versus $\theta$ at $T = 2.0$ for (a) $a = 10^{-4}$, (b) $a = 0.9$, (c) $a = 0.9999$, and (d) $a = 1.0001$. The values of $a$ are the same as in Kerswell et al. [32]. The optimum $\theta$ maximizing the gain is shown by the solid marker, while other samples drawn during the Bayesian optimization are represented by hollow markers. The predicted mean by the constructed GPR is close to the true gain obtained by solving Eq. (11) by the fourth-order Runge-Kutta scheme given the optimum $\theta$. The normalized error between the GPR and true gains measured in $L_2$-norm is equal to 0.0102, 0.0142, 0.1360, and 0.4267 for $a$ equals to $10^{-4}$, 0.9, 0.9999, and 1.0001, respectively. In constructing the GPR, the $\sigma$ in the observational noise in Eq. (2) was considered to be zero. The shaded areas specify the 95% confidence intervals built around the posterior predictive mean.

## 4. Shape optimization of a lid-driven cavity

We now move to problems involving the Navier–Stokes equations. Consider an incompressible two-dimensional cavity flow, where all walls are at rest except the upper wall (lid) which moves with the constant velocity $U_0$ in the positive horizontal direction, $x$. The aim is to optimize the shape of the left and right walls so that the energy dissipation is either minimized or maximized with the constraint of keeping the total volume of the fluid confined in the cavity fixed and also retain the length and position of the upper and lower walls. Therefore, the objective is to find the wall shape $\Omega$ which either minimizes or maximizes,

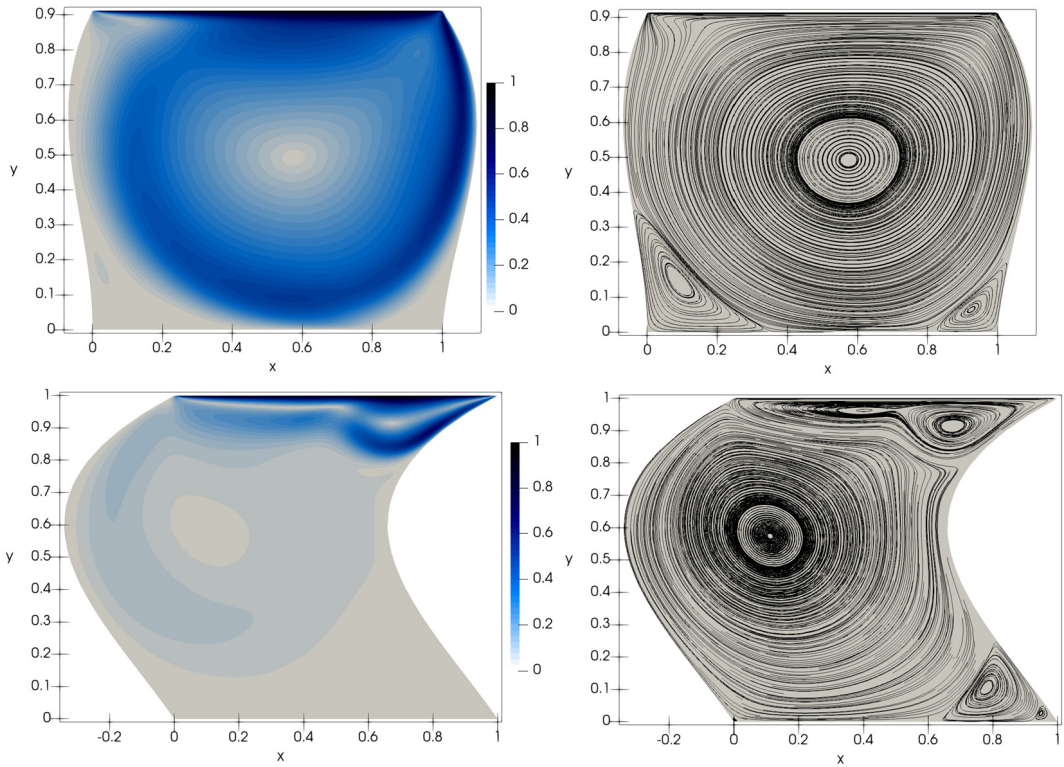

**Fig. 2.** Contours of the velocity magnitude (normalized with the lid velocity $U_0$) and streamlines of the cavity flow with (top) minimum and (bottom) maximum energy dissipation. Third-order polynomials are used to parameterize the side walls of the cavity. The flow is steady with Re = 2000.

$$D = \frac{2}{\text{Re}} \int_\Omega S_{ij} S_{ij} d\mathbf{x}, \qquad (12)$$

keeping $\int_\Omega d\mathbf{x} = m$ fixed. In Eq. (12), **S** is the strain rate tensor, where $S_{ij} = (\partial u_i/\partial x_j + \partial u_j/\partial x_i)/2$ for $i, j = 1, 2$ and $u_i$ denotes the $i$-th component of the velocity vector. In order to apply the Bayesian optimization of Section 2, we parameterize the geometry using a third-order polynomial, $\bar{x} = a_0 + a_1 \bar{y} + a_2 \bar{y}^2 + a_3 \bar{y}^3$, where $\bar{x} = x/l$, $\bar{y} = y/h$, $l$ is the length of the upper and lower walls, $h$ denotes the height of the side walls, and, $x$ and $y$ specify horizontal and vertical coordinates, respectively. Considering the origin of the coordinates at $(x = 0, y = 0)$, the constraint of keeping $l$ fixed leads to, $a_{0,L} = 0$ and $a_{3,L} = -(a_{1,L} + a_{2,L})$ for the left wall, and, $a_{0,R} = 1$ and $a_{3,R} = -(a_{1,R} + a_{2,R})$ for the right wall. As a result, four design parameters are left to optimize, $\mathbf{q} = (a_{1,L}, a_{2,L}, a_{1,R}, a_{2,R})$. The admissible space of $\mathbf{q}$ is the tensor-product of $\mathbb{Q}_{a_{1,L}} = \mathbb{Q}_{a_{1,R}} = [-0.7, 0.7]$ and $\mathbb{Q}_{a_{2,L}} = \mathbb{Q}_{a_{2,R}} = [-0.5, 0.5]$ which are arbitrarily chosen to allow for different wall shapes to occur. For any sample taken from $\mathbb{Q}$, first the shape of side walls is determined. At this stage, the height of the walls, $h$, is adjusted to satisfy the constant-volume constraint. Then, the mesh is generated using gmsh [23]. The simulations are performed using Nek5000, an open-source spectral-element-based flow solver developed by Fischer et al. [19]. In Nek5000, the weak form of the incompressible Navier–Stokes equations are integrated in time on Gauss-Lobatto-Legendre (GLL) points through expressing the velocity and pressure fields in terms of Lagrange interpolants of the Legendre polynomials, see the details in Deville et al. [16]. To avoid the discontinuity in the velocity at the top corners where the moving lid and the side walls meet, the smoothing suggested in Ref. [47] is applied. In all simulations, 30 elements in both $x$ and $y$ directions are considered with 10 GLL points in each spatial directions per element. Simulations start at $t = 0$ from a uniform velocity field and continue until $t = 50 l/U_0$, which is chosen to ensure the steady-state value for dissipation (12) has been reached. For both minimization and maximization problems, the BO starts from same four initial samples for $\mathbf{q}$ that are $[0, 0, 0, 0]$, $[0.7, 0.5, -0.7, 0.5]$, $[0.7, -0.5, -0.7, 0.5]$, and $[0.7, 0.5, -0.7, -0.5]$. Then, decision about next sample of the design parameters is taken. This algorithm is repeated until optimal design parameters are obtained. For this example we only consider steady laminar flow, however an extension to turbulent flow is possible in a straightforward way.

For a cavity flow at Reynolds number Re $= U_0 l/\nu = 2000$ ($\nu$ is the kinematic viscosity), Fig. 2 shows the obtained shapes of the cavity with minimum and maximum energy dissipation. The obtained shapes are consistent with the flow physics, for instance, in case of the maximum dissipation two vortices adjacent to the lid are generated which prevent the fluid underneath to receive a high velocity. The obtained optimal shapes can be compared to results reported by Nakazawa [45] using an adjoint method for optimization, although the setups in the two studies are slightly different. In Ref. [45] the lid velocity was assumed to vary with $x$ (it might be hard to interpret it physically) and the Reynolds number Re was higher than the critical value of about 8000 [6] (although no treatment for turbulence modeling was mentioned). For the case of minimizing the dissipation, the shape in Fig. 2 (top) agrees well with [45]. But we found the shape in Fig. 2 (bottom) has a slightly higher dissipation compared to the maximum-dissipation case reported by Nakazawa [45] in which both side walls were deformed towards the inside of the cavity. This can be an indication for the ability of the Bayesian optimization in exploring different possible geometries and finding the global optimum without quickly getting stuck in a local optimum.

Following the discussion in Section 2 regarding the convergence of the BO, the reported optimal shapes in Fig. 2 are based on the parameters which were found to be the best values within a sample size of $n = 50$. In fact, the parameters for minimizing and maximizing the dissipation were already found at the 20-th and 15-th iterations, respectively. However, the extra samples were taken to ensure the parameter space is sufficiently explored. It is recalled that the definition of the convergence criterion in Bayesian optimization is not straightforward, see [9,66]. Therefore, in order to impose an appropriate stopping criterion, the whole history of the samples has to be considered. The history of the BO-GPR for the optimization cases of the cavity flow is plotted in Figs. A.17 and A.18 in Appendix A.

## 5. Shape optimization of a diverging channel

### 5.1. Motivation

Studying in a controlled fashion pressure-gradient (PG) turbulent boundary layers (TBLs) is important since they are present in a wide range of industrial applications, for instance, the flow around the curved boundaries such as car bodies, airplanes, airfoils, etc. The so-called Clauser pressure-gradient parameter $\beta$ [12,13] is widely accepted as one of the most relevant non-dimensional parameters for PG TBLs [64], which is defined as,

$$\beta = \frac{\delta^*}{\tau_w} \frac{\mathrm{d}P_e}{\mathrm{d}x}, \tag{13}$$

where $\delta^*$, $\tau_w$ and $\mathrm{d}P_e/\mathrm{d}x$ are, respectively, the displacement thickness, the magnitude of the wall-shear stress and the pressure gradient along the boundary-layer edge in the streamwise direction. Since the history of the PG $\beta(x)$ has a significant impact on the characteristic of the TBLs, flows with a constant $\beta$ in the streamwise direction are very relevant for the study of PG TBLs [5,18]. Despite their importance, there are few studies in the literature on flows with constant pressure gradient, mainly due to the difficulty of setting up this configuration with sufficient accuracy. In experiments, a desired $\beta$ distribution at the edge of a TBL is usually achieved by changing the wall shape of wind tunnels through a trial-and-error process, see *e.g.* Sanmiguel Vila et al. [64]. This is, in fact, very time consuming, since $\beta$ is needed to be measured in the streamwise direction at each iteration. Consequently, it is very difficult to achieve a long constant-$\beta$ region in the streamwise direction. If a proper shape of the wall could be known prior to an experiment, not only time and cost could be saved, but also a more accurate constant-$\beta$ distribution would be eventually obtained.

The objective of this section is to efficiently obtain the optimal shape of the upper wall of a channel using the Bayesian optimization described in Section 2 so that a target $\beta$ along the edge of the TBL over the lower wall is achieved. We consider both zero-pressure-gradient (ZPG) and adverse-pressure-gradient (APG) TBLs with constant values for the target $\beta$. Besides these, an optimization is considered where the target $\beta$ distribution varies in the streamwise direction. Such distribution is taken from the flow around a NACA0012 airfoil. This demonstrates the applicability of the BO method to a wide variety of target distribution for the pressure gradient $\beta$.

### 5.2. Problem setup

The considered flows have high Reynolds number and are hence turbulent. Therefore, we perform incompressible Reynolds-averaged Navier–Stokes (RANS) simulations using the open-source software OpenFOAM [83]. In particular, the `simpleFoam` solver is used which is based on the SIMPLE scheme [55] for velocity-pressure coupling. The $k-\omega$ shear-stress transport (SST) turbulence model introduced by Menter et al. [39] is used, since for TBLs this model has shown better agreement with experimental data compared to other models, see *e.g.* Vinuesa et al. [79]. Note that the default values are used for the RANS model coefficients.

A two-dimensional domain is considered to mimic the conditions that may exist in a wide wind tunnel, see the schematic in Fig. 3. The domain has a fixed-shape flat-plate lower wall and an upper wall whose shape is subject to optimization. The domain length in the streamwise direction and the domain height at the inlet are denoted by $L_x$ and $L_y^{\mathrm{in}}$, respectively. In particular, we consider $L_x/\delta_{99}^{\mathrm{in}} = 1000$ and $L_y^{\mathrm{in}}/\delta_{99}^{\mathrm{in}} = 40$, where $\delta_{99}^{\mathrm{in}}$ is the 99% boundary layer thickness at the inlet. In general, $\delta_{99}$ is defined as the distance from the wall at which the mean streamwise velocity becomes 99% of the local edge velocity, $U_e$. In the present study, $U_e$ represents the local maximum velocity of the lower-wall's TBL. Note that the diagnostic-plot scaling method proposed by Vinuesa et al. [80] is not applicable to RANS due to the lack of information about the instantaneous velocity. Note also that $\delta_{99}^{\mathrm{in}}$ pertaining to the upper and lower TBLs are the same since the inflow condition for the ZPG-TBLs over the upper and lower walls are taken to be the same. The inflow profiles are assigned based on the DNS data of ZPG-TBL from Schlatter and Örlü [65], see the details in Ref. [43]. The upper wall is constructed by using a spline function to smoothly connect the intersection of the channel inlet and the upper wall, *i.e.* $(x = 0, \ y = L_y^{\mathrm{in}})$, and $d$ control points. The control points are equally spaced in the streamwise direction. The heights $y$ of these points are subject to optimization, therefore the $(x, y)$ coordinates of the $i$-th control point are defined as $(iL_x/d, q_i)$ where $i = 1, 2, \cdots, d$, see Fig. 3. When applying the BO of Section 2, for each sample of $\mathbf{q}$ a new geometry for the domain is obtained for which a structured hexahedral mesh is generated using OpenFOAM standard meshing function, `blockMesh`. In order to resolve the near-wall region of the TBL, a fine mesh resolution is needed near the wall. The mesh is constructed in such a way that $y_1^+ = u_\tau y_1/\nu$ remains below unity at the walls. Here, $y_1$ is the distance of the first computational cell center from the lower wall, $\nu$ is the kinematic viscosity, $u_\tau = \sqrt{\tau_w/\rho}$ is the wall-friction velocity with $\tau_w$ and $\rho$ respectively denoting the wall-shear stress and fluid density.

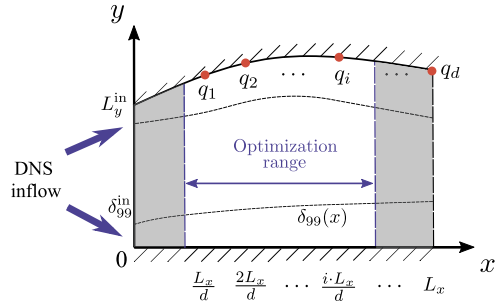

**Fig. 3.** Schematic of the domain used in the shape optimization of the diverging channel. The black dashed lines illustrate the TBLs over the upper and lower walls. Red circles denote the control points, whose heights $q_i$ are to be optimized. The optimization range is set to avoid any inlet or outlet effects on the TBLs. (For interpretation of the colors in the figure(s), the reader is referred to the web version of this article.)

**Table 1**
Summary of the three optimization cases. The admissible space $\mathbb{Q}$ is non-dimensionalized by $\delta_{99}^{in}$, i.e. $\tilde{\mathbb{Q}} = \mathbb{Q}/\delta_{99}^{in}$. The first sample $\mathbf{q}^{(1)}$ is randomly taken within the corresponding admissible range $\mathbb{Q}$. The total number of iterations is denoted by $n$.

| Case name | Constant-0 | Constant-1 | Wing | |
|---|---|---|---|---|
| $\beta_t$ | 0 | 1 | NACA0012 | |
| $\tilde{\mathbb{Q}}$ | $\tilde{\mathbb{Q}}_1 = [40, 44]$<br>$\tilde{\mathbb{Q}}_2 = [40, 46]$ | $\tilde{\mathbb{Q}}_1 = [55, 65]$<br>$\tilde{\mathbb{Q}}_2 = [75, 85]$<br>$\tilde{\mathbb{Q}}_3 = [85, 100]$<br>$\tilde{\mathbb{Q}}_4 = [95, 110]$ | $\tilde{\mathbb{Q}}_1 = [40, 42]$<br>$\tilde{\mathbb{Q}}_3 = [40, 45]$<br>$\tilde{\mathbb{Q}}_5 = [40, 50]$<br>$\tilde{\mathbb{Q}}_7 = [45, 60]$ | $\tilde{\mathbb{Q}}_2 = [40, 45]$<br>$\tilde{\mathbb{Q}}_4 = [40, 50]$<br>$\tilde{\mathbb{Q}}_6 = [40, 60]$<br>$\tilde{\mathbb{Q}}_8 = [50, 80]$ |
| $\mathbf{q}^{(1)}/\delta_{99}^{in}$ | [41.3, 43.1] | [63.2, 79.4, 88.5, 103.0] | [40.4, 40.9, 44.2, 42.1, 42.0, 52.3, 58.9, 69.0] | |
| Optimization range | $200 < x/\delta_{99}^{in} < 900$ | $300 < x/\delta_{99}^{in} < 900$ | $150 < x/\delta_{99}^{in} < 950$ | |
| $n$ | 25 | 50 | 80 | |

The aim of the optimization is to obtain a pressure-gradient distribution sufficiently close to a given target $\beta_t$. Therefore, we can formulate a minimization problem whose objective function is defined as an error between the computed $\beta$ distribution and the target distribution,

$$r(\mathbf{q}) = \|\beta(\mathbf{q}, x) - \beta_t(x)\|_{L_2}, \tag{14}$$

where, $\|\cdot\|_{L_2}$ denotes a standard $L_2$-Lebesgue norm evaluated over the domain of $x$. Note that regions close to the inlet and outlet of the domain have to be excluded when evaluating $r(\mathbf{q})$ to avoid any inlet or outlet effects on the $\beta$ distributions, see Fig. 3. The remaining range of $x$ over which the norm in Eq. (14) is computed is referred to as optimization range which may be chosen adequately for each optimization case as explained in Section 5.3. For additional details about the problem setup, refer to Ref. [43].

### 5.3. Optimization

As explained in the Section 5.1, optimization is conducted for three different distributions of $\beta$: constant values of zero (ZPG) and one (APG), and a non-constant distribution corresponding to the flow around a NACA0012 airfoil. These optimization cases are named as Constant-0, Constant-1 and Wing, respectively, and are discussed in the following subsections. The conditions of these cases are summarized in Table 1. Prior to the optimization, numerical experiments are conducted to decide the number of control points, admissible range of parameters, optimization range, and the total number of iterations.

#### 5.3.1. Constant-$\beta$ distribution

For the Constant-0 case, essentially a correction for the displacement effect of the growing boundary layers is expected, which yields a comparably simple shape of the upper wall. Thus we chose to only use 2 points along the length of the channel. The optimization of the Constant-0 case is conducted and the optimum is found at the 22-nd iteration. Note that the optimum is taken as the best value over the first $n = 25$ iterations and the history of the optimization is plotted in Figs. A.19 and A.20 in Appendix A. Due to the definition of the objective function, see Eq. (14), the validity of the computed optimal parameters can be examined. The results of the optimization are shown in Fig. 4. It is observed that the global minimum of $r$ is found in the considered admissible space and the associated $\beta$ distribution is highly accurate (within $\pm 0.005$ of the target value of zero). The 95% confidence interval of $r(\mathbf{q})$, which is shown in Fig. 4(d), is smaller close to the optimum because of the higher number of samples.

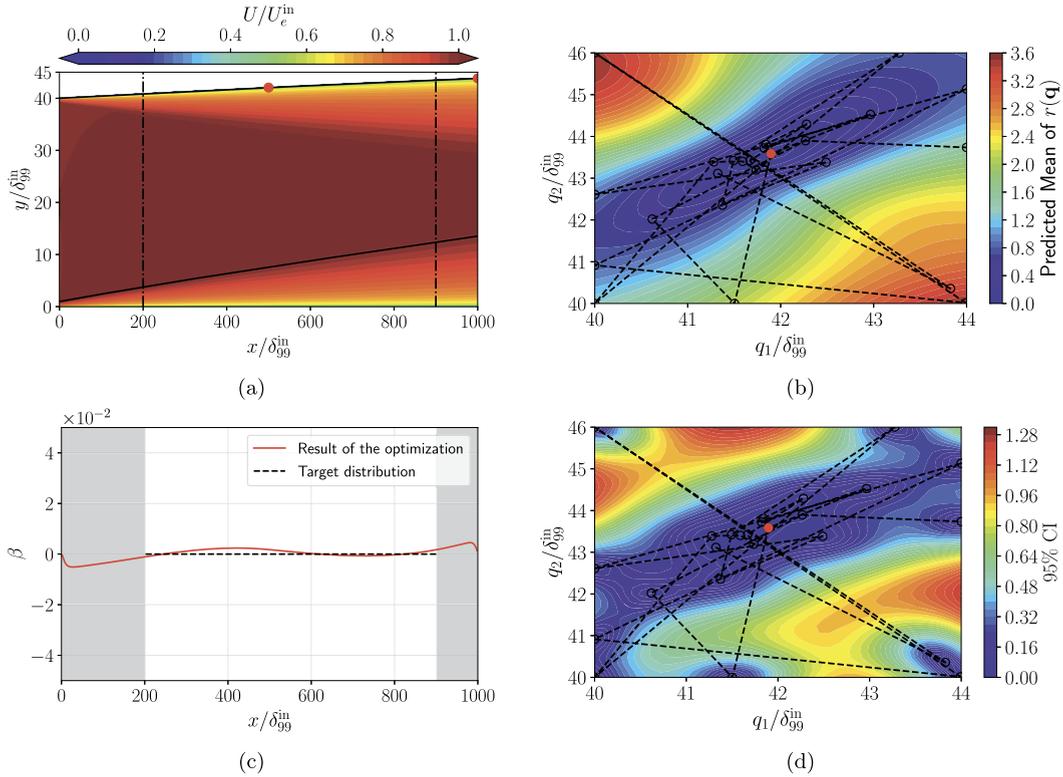

**Fig. 4.** Results of the optimization of the Constant-0 case using 2 control points: (a) contours of the streamwise velocity, (b) contours of the mean prediction of the objective function (14), (c) comparison of the computed $\beta$ distribution with the target value, (d) contours of the 95% confidence interval of the objective function. The red markers show the optimal parameter $\mathbf{q}_{\text{opt}}/\delta_{99}^{\text{in}} = (42.05, 43.80)$. The vertical black dash-dotted lines in (a) and the gray region in (b) illustrate the range excluded from the optimization. The black solid line in (a) shows $\delta_{99}$ of the TBL over the lower wall. The black hollow markers and dashed lines in (b) and (d) show the sampled parameters and their trace, respectively. Note that the streamwise velocity is non-dimensionalized by the edge velocity at the inlet, $U_e^{\text{in}}$.

Similar to the Constant-0 case, optimization of the Constant-1 case, i.e. for $\beta_t = 1$ is conducted. Since the target distribution of $\beta$ is constant but larger than zero, the optimum shape of the upper wall is expected to be only moderately complex; thus, 4 control points are used. The optimum is found at the 40-th iteration, whose results are shown in Fig. 5. The optimal shape of the upper wall is steeper than that of the Constant-0 case, since the target $\beta_t$ is higher. In the Constant-1 case, the computed optimal $\beta$ distribution remains almost constant in the optimization range and deviates from the target $\beta_t$ by less than $\pm 0.02$. It is noteworthy that since the inflow is taken from a ZPG-TBL, a development length is needed to reach $\beta = 1$. This development length is chosen to be $300\delta_{99}^{\text{in}}$. We observed that forcing a faster development of the $\beta$ through imposing a shorter development length, could lead to the separation of the TBL over the upper wall and hence make it more difficult to achieve a constant $\beta$ for the lower-wall's TBL. As separation is typically unwanted in a wind-tunnel setting, we prefer to have a longer development length avoiding this issue.

To validate the results of the constant-$\beta$ TBLs obtained from the optimization, the streamwise development of the skin-friction coefficient $c_f$ and the shape factor $H$ are compared to the reference data, see Fig. 6. The skin-friction coefficient and the shape factor are defined as $c_f = 2(u_\tau/U_e)^2$ and $H = \delta^*/\theta$, where $\delta^*$ and $\theta$ respectively denote the displacement and momentum thickness. As the reference data for the constant-$\beta$ TBLs, the correlations based on the data compiled by Vinuesa et al. [81] and the experimental data from Sanmiguel Vila et al. [64] are adopted. The Reynolds number based on the momentum thickness, $\text{Re}_\theta = U_e\theta/\nu$, is used as the horizontal axis of the plots in Fig. 6, so that the direct comparison is possible. Clearly, the optimized cases show excellent agreement with the reference data in the region where $\beta$ is approximately constant ($\text{Re}_\theta \gtrsim 3000$). The Constant-1 case shows a transition from $\beta = 0$ to $\beta = 1$ profile because of the development of $\beta$ in the streamwise direction, see Fig. 5(b). Note that the experimental $\beta$ distribution of Sanmiguel Vila et al. [64] has a variation in the streamwise direction (about ±15%) because of the difficulties to set the constant-$\beta$ distributions in the experiments, as mentioned in Section 5.1. As a conclusion, not only the applicability of the optimization method to obtain constant-$\beta$ distributions is confirmed but also the validity of the resulting constant-$\beta$ TBLs is approved.

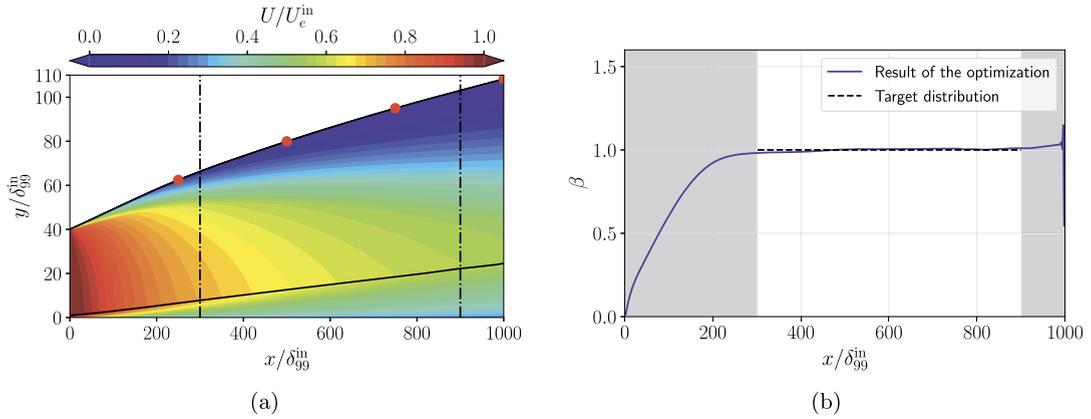

**Fig. 5.** Results of the optimized flow in Constant-1 case using 4 control points: (a) contours of the streamwise velocity, (b) comparison of the computed $\beta$ distribution with the target value. The optimal parameter is $\mathbf{q}_{opt}/\delta_{99}^{in} = (62.5, 79.9, 95.0, 108.0)$. The descriptions in the figure are the same as Fig. 4. Note that the range of the vertical axis in Fig. 4(a) and Fig. 5(a) are set differently for illustration purposes.

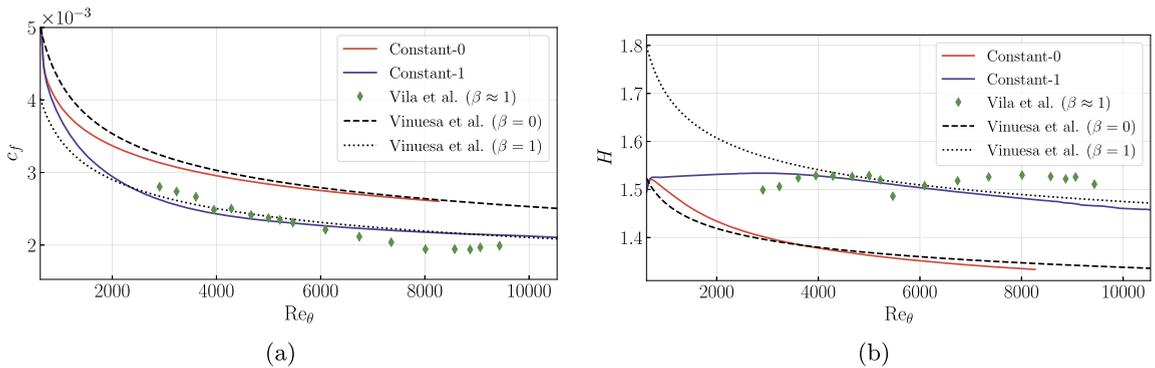

**Fig. 6.** Validation of the optimized Constant-0 and Constant-1 cases: (a) streamwise development of the skin-friction coefficient $c_f$, (b) streamwise development of the shape factor $H$. Reference correlations and experimental data are based on the data compiled by Vinuesa et al. [81] and Sanmiguel Vila et al. [64], respectively.

#### 5.3.2. Non-constant-$\beta$ distribution

The same optimization procedure can be applied when $\beta_t$ is not constant in the streamwise direction. The target $\beta$ distribution is taken from the flow around a NACA0012 airfoil. Note that due to the geometrical symmetry with respect to the airfoil chord, $\beta$ distributions of the suction and pressure sides are identical for NACA0012 at zero angle of attack. The distribution of the target $\beta$ is taken from the numerical simulation by Tanarro et al. [74]. Since the target $\beta$ is originally defined based on the normalized coordinate $x/c$, where $c$ denotes the chord length, it has to be mapped to the channel normalized coordinate in the streamwise direction, i.e. $x/\delta_{99}^{in}$. Here, the original data are mapped to the first 95% of the channel length, i.e. $0 < x/c < 1 \rightarrow 0 < x/\delta_{99}^{in} < 950$, to avoid any effect of the outlet boundary. Furthermore, the first 15% of the channel is excluded from the optimization range because the reference data close to the leading edge is not available. According to Ref. [74], this is due to the difficulty of calculating $\delta_{99}$ with the diagnostic-plot scaling method near the leading edge since the flow is not fully turbulent. To be able to achieve the relatively complex $\beta$ distribution along the edge of the TBL at the lower wall, the parameterization of the upper wall of the channel is done with 8 control points, see Table 1.

The results of the optimization are shown in Fig. 7. The optimum is found at the 52-nd iteration with associated $\beta$ being found very close to the target distribution. In fact, the largest deviation which occurs around $x/\delta_{99}^{in} \approx 850$ is less than 0.54. This promising result proves that the Bayesian optimization method can also be used for more complex pressure-gradient distributions. This is, in fact, very useful for future wind-tunnel experiments to study wall-bounded turbulent flows with arbitrary $\beta$ distributions, noting that the common approach in practice is based on trial-and-error, which can be inaccurate and time consuming.

The results of the present section have been obtained by RANS, which may have its own deficiency in fidelity compared to e.g. large eddy simulations or even direct numerical simulations. However, the underlying flow simulation method does not affect the performance of the optimization procedure. Therefore, the current results could be even achieved using higher-fidelity methods, or even – if properly implemented – in an experimental setup.

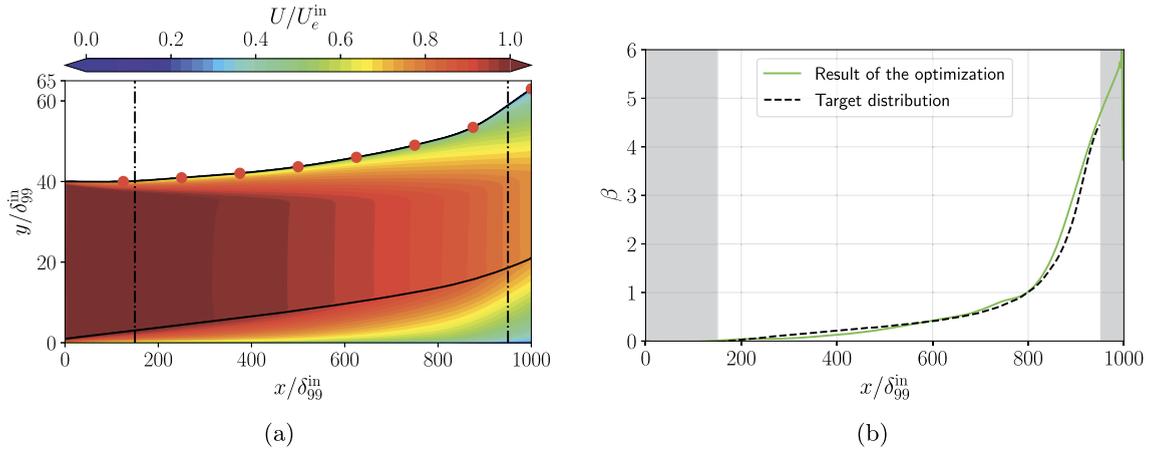

**Fig. 7.** Results of the optimized flow in Wing case using 8 control points: (a) contours of the streamwise velocity, (b) comparison of the computed $\beta$ distribution with the target $\beta$ distribution. The target distribution is taken from the simulation data by Tanarro et al. [74]. The optimal parameter values are $\mathbf{q}_{opt}/\delta_{99}^{in} = (40.0, 40.9, 42.0, 43.7, 46.0, 49.0, 53.4, 63.0)$. The descriptions in the figure are the same as Figs. 4 and 5.

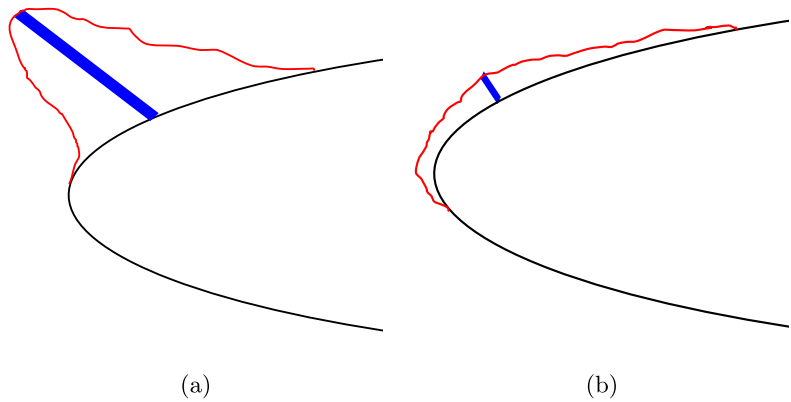

**Fig. 8.** The schematics of the horn ice (a) and streamwise ice (b). The black, red and blue lines respectively illustrate the surface of a clean airfoil, the actual ice shape, and an example of the spoiler-ice configuration employed to model the aerodynamic effects of the actual ice.

## 6. Optimization of a spoiler-ice airfoil model

Finally, in this section we present a more applied flow case which shows the potential of the BO framework also for industrial flow cases where optimal designs are sought.

### 6.1. Motivation

When operating in cold climate, the performance of the wind turbines can be reduced by icing [8,72]. In an extreme condition, heavy icing can even lead to a complete stop of the turbine. Icing is also a risky condition for airplane wings since it may induce flow separation at a small angles of attack, which then might lead to stall and consequently loss of lift. Ice accretion is a complex process which depends on both aerodynamics and thermodynamics. The process is affected by many parameters, for instance, ambient temperature, surface properties, relative speed of the airfoil, and the time span over which the icing event takes place. As a result, numerous possibilities for an ice profile exist. Bragg et al. [8] categorized the leading-edge ice on the airfoils into four types: roughness, horn, streamwise, and spanwise-ridge ice. In this study, we focus on the horn and streamwise ice, see Fig. 8.

The main characteristic of the horn ice is the presence of the large protuberances which induce a large flow separation downstream of the ice, and hence dramatically affect the aerodynamic performance, including the lift, drag and moment coefficients. On the other hand, streamwise ice is smoothly formed along the streamlines and is less dangerous in the sense of having adverse effects on the aerodynamic performance.

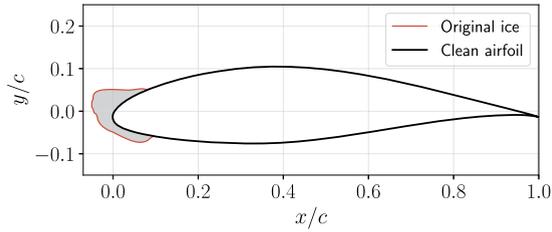

**Fig. 9.** Original iced airfoil used in the present study. The ice is formed on a clean NACA64618 airfoil, which is shown with a black solid line. The original ice shape is illustrated with a red solid line and a filled gray region. Coordinates are non-dimensionalized by the chord length of the clean airfoil, *c*. Note that the origin of the *x* coordinate is set as the leading edge of the clean airfoil, and this is the convention throughout Section 6.

Modeling an iced airfoil is a challenging task in both CFD and experiments since the ice profile essentially has infinite degrees of freedom, noting that every ice shape is unique and consequently the investigations would be case-dependent. In order to generalize the icing assessments, one needs to find specific characteristics to parameterize the geometry such that various ice profiles can be described with the reduced-order model. The simplified "simulated-ice" concept has been considered as a representative of the actual ice profile. In a simple representation, Papadakis et al. [50–54] approximates the horn-shaped ice from glaze ice conditions as spoilers. As a result, the profile of the horn is reduced to a thin plate and its effect on the aerodynamic properties is parameterized by just its height, angle, and location. In particular, the thin plates which are called the "spoilers" and have the same height as the original ice can be used to reproduce the aerodynamic performance of the ice, see Fig. 8. The resulting model is called the "spoiler-ice" model. Although the height, angle, and location of the spoilers have been considered as important parameters which affect the airfoil aerodynamic performance, see *e.g.* Refs. [52,51], Tabatabaei et al. [70,71] suggested that the thickness of the spoilers should be also taken into account as an effective parameter.

Although parametric studies of the spoiler-ice model have been conducted to investigate associated effects on the aerodynamic performance, see *e.g.* Refs. [52,51], it is not yet clearly known which configuration of the spoiler is appropriate to accurately represent the aerodynamic performance of an arbitrary given real ice profile. The BO method introduced in Section 2 can be utilized to find the optimal parameters of the spoiler configuration. Although the spoiler-ice concept has been previously used mainly for the horn ice, see *e.g.* Ref. [53], here we also consider to apply the idea to the streamwise ice. Therefore, an original arbitrary ice shape is chosen so that it has both horn and streamwise ice on the suction and pressure side of the airfoil, respectively. Two spoilers on the suction and pressure sides (the upper and lower spoilers, respectively, shown in Fig. 10(a)) are used to obtain a similar effect of the original iced airfoil. Since a spoiler-ice with the same height as the original ice shape has been used in the previous studies [51–53], here the optimization is first conducted with the constraint of keeping the heights of the spoilers fixed. However, it is shown in Fig. 14 that the results of the optimization without such a constraint are in better agreement with those of the original ice. It will thus be shown that a more complete parameter optimization may indeed be relevant for accurate simulations.

*6.2. Problem setup*

In this section, we describe the setup and implementation of the icing flow case, including the basic simulation, the meshing and optimization loop.

*6.2.1. Original ice profiles and the spoiler ice*

As clean airfoil, a NACA64618 profile is considered, which has been used in a 5MW NREL (National Renewable Energy Laboratory) wind-turbine blade. For the purpose of the optimization, an arbitrary original ice shape is required as a reference which is taken from the ice formed on the leading edge of the airfoil studied in Ref. [22], see Fig. 9. The ice on the upper (suction) and lower (pressure) sides of the airfoil is of the horn and streamwise types, respectively.

As mentioned in Section 6.1, the flow around the original iced airfoil can be modeled by two spoilers each located on one side of the airfoil. For the purpose of optimization, the configuration of the spoilers is controlled by seven parameters: heights, angles and widths of the upper and lower spoilers, $h_u$, $h_l$, $\theta_u$, $\theta_l$, $w_u$ and $w_l$, respectively, and the displacement of the lower spoiler, ($d_l$), see Fig. 10(a). The displacement of the upper spoiler is not taken into account since the original ice shape on the suction side is horn ice, which has a clear connection between the position of the horn of the original ice and the position of the spoiler-ice. On the other hand, the pressure side of the original ice is of streamwise type, hence it is not, in general, clear where the spoiler-ice should be located. To rectify this, the displacement parameter $d_l$ is needed for the lower spoiler to optimize its position as well as the other parameters.

Prior to the optimization, several numerical experiments are conducted to decide the admissible range of the parameters and the total number of iterations. Fig. 10(b) shows the configuration of the spoilers when the design parameters are at either end of their admissible range. Note that the upper limit of the admissible range of the heights, $h_u$ and $h_l$, is the same as the highest point of the original ice profile.

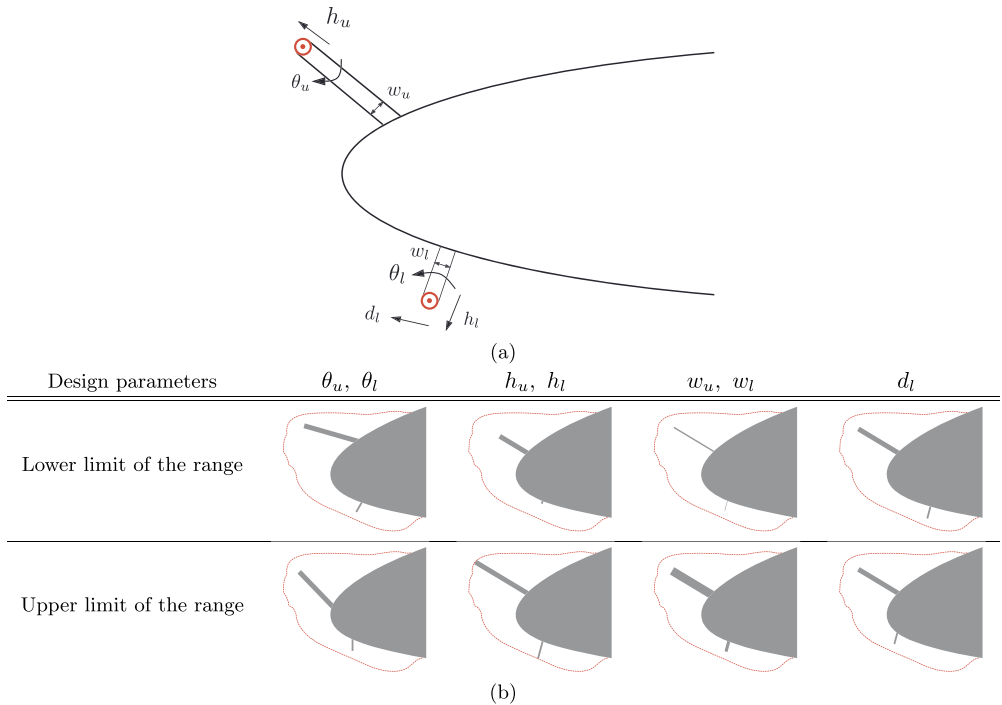

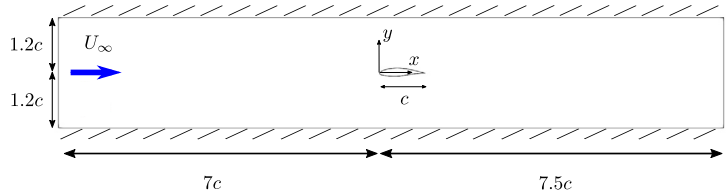

**Fig. 10.** Schematic representation of the spoiler ice on the airfoil to model the original ice profile: (a) design parameters, (b) configuration when the design parameters are at the either end of their admissible range. The red circles in (a) specify the rotation center of $\theta_l$ and $\theta_u$. The direction of $d_l$ is defined to be parallel to the tangential direction at the root of the lower spoiler. The red dashed line in (b) shows the original ice profile. Note that in the particular configuration at (b) all parameters are fixed at the middle of their admissible range except the specified design parameters.

**Fig. 11.** Schematic of the domain used in the iced-airfoil simulations. The chord length of the airfoil and the freestream velocity are denoted by $c$ and $U_\infty$, respectively.

*6.2.2. Numerical simulations*

Numerical simulations are conducted considering a cross section of an iced-airfoil inside a (virtual) wind tunnel. Two-dimensional steady and incompressible RANS are performed using OpenFOAM with the same solver settings and turbulence model as mentioned in Section 5.2. The computational domain, see Fig. 11, is made based on the test section of the Minimum-Turbulence-Level (MTL) wind tunnel, which is a closed-loop circuit tunnel housed at KTH. Additional information about the MTL wind tunnel can be found in *e.g.* Ref. [37].

A uniform inflow condition for the velocity is applied at the inlet of the domain. The resulting Reynolds number based on the airfoil chord is $Re_c = U_\infty c/\nu = 4.0 \times 10^5$, where $U_\infty$ and $c$ respectively denote the freestream velocity and the chord length of the airfoil. No-slip/no-penetration boundary conditions are imposed at the top and bottom walls of the domain, as well as on the airfoil, ice, and the spoiler ice.

The commercial software Ansys ICEM CFD 18.2 [1] is used for generating high-quality meshes in the relatively complex geometries involved in the problem. The mesh is constructed to be sufficiently fine, and in particular $y_1^+$ is kept below 1 all around the airfoil. The resulting total number of computational cells is about 1 million. Note that the number of computational cells slightly differs for the simulations associated with a particular configuration of the spoiler ice. To put the computational cost in perspective, each flow simulation takes about one hour running in parallel on 24 cores of Intel Xeon E5-2690v3 Haswell processors.

To measure the similarity of the aerodynamic performance of a spoiler-ice model and the actual ice profile, the pressure coefficient ($c_p$) distribution around the airfoil is considered, since it is directly related to the aerodynamic characteristics such as the lift and drag coefficients. The objective function in the optimization is thus defined as the $L_2$-norm of the error between the computed $c_p$ distributions on the suction and pressure sides of the airfoil, and the corresponding distribution of the original iced airfoil. The pressure coefficient is defined as $c_p = 2(p - p_\infty)/(\rho U_\infty^2)$, where $p$ and $p_\infty$ denote the local pressure on the surface of the airfoil and the freestream pressure at the inlet, respectively. The objective function reads,

$$r(\mathbf{q}) = \Sigma_{f=1}^2 \sqrt{\int_{0.1c}^{c} |c_{p_f}(\mathbf{q}, x) - c_{p_f,t}(x)|^2 \, dx}, \tag{15}$$

where $\Sigma_{f=1}^2$ specifies the summation over the suction and pressure sides of the airfoil and $c_{p,t}$ denotes the $c_p$ distribution of the original iced airfoil. Note that the objective function is evaluated only downstream of the original ice, meaning that the optimization range is set to $0.1 < x/c < 1$ for both suction and pressure sides of the airfoil.

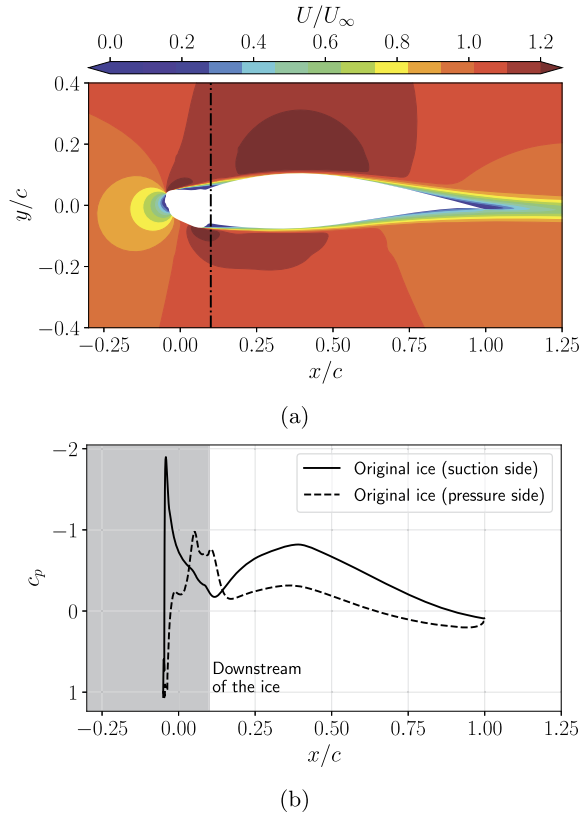

(a)

(b)

**Fig. 12.** Results of the original iced-airfoil case: (a) contours of the streamwise velocity, (b) the pressure-coefficient distribution. The velocity is non-dimensionalized by the freestream velocity at the inlet. The black dash-dot line in (a) shows the beginning of the optimization range. The gray area in (b) is the range excluded from the optimization. Note that the origin of the $x$ coordinate is set at the leading edge of the clean airfoil, so the iced airfoil starts from negative $x/c$, see Fig. 9. The range of the horizontal axes are chosen to be the same so that comparison between two figures becomes easier.

Prior to the optimization, a simulation of the original iced airfoil is conducted to obtain the target pressure-coefficient distribution $c_{p,t}$. The resulting velocity contours and pressure-coefficient distribution are shown in Fig. 12. Because of the ice formed on the leading edge, there are separation bubbles right after the ice on both suction and pressure sides. Moreover, the $c_p$ near the leading edge exhibits sudden changes in the streamwise direction. Note that $x/c \lesssim 0.1$ is still on the ice surface (in fact, the ice ends at $x/c = 0.086$ and $0.0975$ on the suction and pressure sides, respectively). Therefore, the optimization range is set to $0.1 < x/c < 1$, as represented in Fig. 12.

### 6.3. Optimization

Optimization of the design parameters is conducted with two configurations: spoilers with fixed height and flexible height. As mentioned in Section 6.1, glaze ice has been simulated by spoiler ice whose height is the same as the highest point of the original ice [51–53]. As a result, one optimization is conducted with the constraints of having the heights of the spoilers fixed, and the case is referred to as "fixed-height optimization". Imposed by these constraints, the number of design parameters is reduced to $d = 5$. Since the height of the spoiler ice is a relevant parameter on the aerodynamic performance [72,15], the second optimization is conducted without the above constraints, and the case is called "flexible-height optimization". Allowing the height of the spoilers to change, the number of design parameters becomes $d = 7$. The maximum number of iterations in the BO-GPR algorithm of Section 2 is chosen as $n = 50$ and 90 for fixed and flexible-height optimizations, respectively. These limits are chosen based on the numerical experiments conducted prior to the optimization. The first sample is taken randomly within the admissible range. The optimal parameters are found for the 13-th and the 43-rd iterations for the fixed and flexible-height optimizations, respectively. The flow fields and $c_p$ distributions of the optimal configurations are compared to those of the original iced airfoil in Figs. 13 and 14, in order to validate the results of the optimizations.

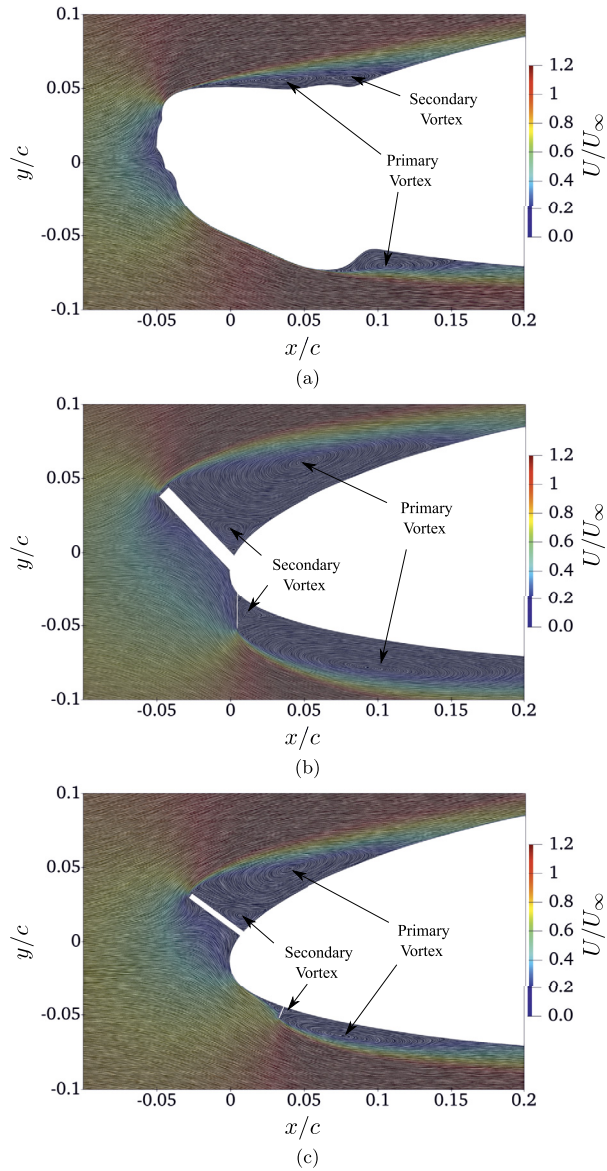

**Fig. 13.** Comparison of the flow structures around the leading edge. The line integral convolution (LIC) technique [10] has been used to visualize the velocity field along with the contours of the streamwise velocity for (a) the original iced airfoil, (b) the optimized spoiler ice with fixed heights, and (c) the optimized spoiler ice with flexible heights.

In Fig. 13, the separation bubbles and vortex structures are illustrated as they appear behind the ice profile and the spoiler ice on both pressure and suction sides. The general structure is in agreement with the previous studies [51,71], where the primary and secondary vortices were observed around the spoiler-ice through numerical simulations. In fact, it is not even necessary that the flows downstream of the original ice and the spoiler ice have similar vortex structures since the primary purpose of using the spoiler-ice is to model the aerodynamic characteristics on an iced airfoil. The flexible-height optimization successfully yields a reattachment point similar to the original iced-airfoil case, while the reattachment point is pushed further downstream for the fixed-height optimization. This has a direct influence on the $c_p$ distribution, see Fig. 14. In fact, the $c_p$ distribution of the fixed-height case has four times larger deviation from the target distribution, *i.e.* $r$ in Eq. (15), compared to the flexible-height case. It is also observed that for the fixed-height optimization, the optimal parameters $\mathbf{q}_{opt}$ are located at the borders of the considered admissible space. This means that the global optimum could change value if the admissible space would be different. Recall that the admissible range of the parameters are limited due to the geometrical and meshing constraints. In contrast, for the flexible-height optimization, the optimum $\mathbf{q}_{opt}$ is found within the admissible space and leads to a better agreement with the result of the original iced airfoil. Therefore, despite being used in the literature, keeping the heights in the spoiler-ice model fixed is not a sufficiently accurate assumption, and considering also the height as an design parameter leads to higher fidelity in the results.

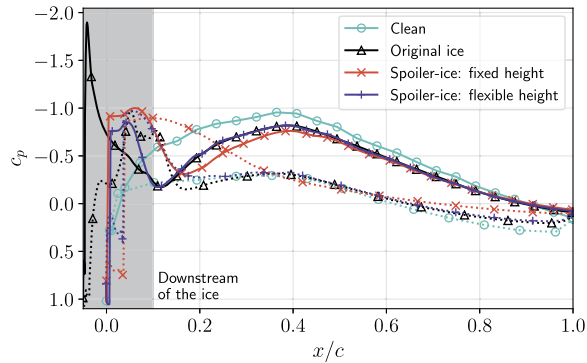

**Fig. 14.** Comparison between the distributions of the pressure coefficient obtained for the original ice profile (shown in Fig. 12), and the optimized spoiler-ice models. The corresponding flow fields are represented in Fig. 13. The solid and dotted lines specify the suction and pressure sides of the airfoil, respectively. The black lines represent the target distribution, $c_{p,t}$, which is also shown in Fig. 12(b). Note that the optimization range is $0.1 < x/c < 1$, which specifies a region downstream of the original ice. The distribution of the clean airfoil is also presented for reference.

The obtained $c_p$ distribution from the flexible-height optimization is almost identical to the target distribution within the optimization range except on the pressure side around $x/c = 0.2$. The slight deviations are due to the fact that the original ice on the pressure side is of the streamwise type, which is basically more difficult to be represented by a spoiler-ice model.

The BO has also shown to be an efficient optimizer even for this applied CFD case. The coupling to the CFD code and meshing software are quite straightforward, and can be done non-intrusively. Note that no derivatives of the target function are needed, which makes the present approach suitable for general CFD applications.

## 7. Summary and conclusions

The Bayesian optimization based on Gaussian process regression (BO-GPR) has been applied to different CFD problems ranging from purely academic to industrially relevant setups, using state-of-the-art simulation methods. The diversity of the examples with different numbers of design parameters helps to better understand the applicability and performance of the BO-GPR approach which has not been frequently used in CFD and turbulent flows despite being a primary choice in other fields, *e.g.* data sciences. The use of the BO-GPR is straightforward noting that the Bayesian optimization is among the gradient-free approaches and hence only requires forward evaluation of the quantities of interest (QoIs), *i.e.* running a CFD code. In the BO, a sequentially-updating set of samples is taken from the space of the design parameters, such that the sequence converges to the global optimum after a finite number of iterations. A main advantage of the BO-GPR is its versatility, meaning the algorithm is non-intrusive and can be utilized with any CFD code and for any number of design parameters. The use of a GPR surrogate provides the possibility of estimating the confidence in the values of the cost function over the admissible space of the parameters. What is needed to create the optimization loop is to automate the whole process through appropriate script-driven interfaces between the BO and CFD codes. The software developed in the present work is based on `GPy` [25] and `GPyOpt` [75] libraries and is available online.[1]

Although the optimization problems studied in the present study contain as many as 8 design parameters, the maximum number of iterations to ensure obtaining a reliable global optimum is no more than 90. This number should be compared with the number of function evaluations necessary *e.g.* for adjoint-based optimizations. More interestingly, by changing the number of design parameters from 2 to 8, the number of required flow simulations does not significantly increase. Therefore, the computational effort for BO-GPR is observed to be affordable for many practical applications, especially if the RANS (Reynolds-averaged Navier–Stokes) approach is employed for the simulation of turbulence. However, similar to other gradient-free approaches, the BO-GPR may become inefficient for a large number of design parameters. As shown for instance by Wu et al. [84], the efficiency of BO-GPR can be improved by adding sensitivity information from the gradients of the cost function computed by adjoint methods. Application and analysis of such modifications will be considered in the future extensions of the present study.

We demonstrate the BO-GPR approach on two comparably simple problems in Sections 3 and 4 for an analytical test problem and a lid-driven cavity. In both cases, our approach could find the global optimum, as compared to previous liter-ature results of the same cases. The optimization required about 20 iterations for the four-dimensional problem, exploring the complete parameter space in an efficient manner. It can be observed that the Bayesian technique quickly focuses on the regions where the potential optima are located. In addition, the uncertainty information about the optima can be retrieved from the approach.

---

[1] https://github.com/KTH-Nek5000/BO_GP.

A more physically relevant problem is considered on the example of the shape optimization of the upper wall of a two-dimensional channel. Here, the goal is to obtain a given target pressure-gradient (PG) distributions in a turbulent boundary layer developing over a flat wall, which is achieved by adjusting the location of the upper wall in a variable-height channel flow. The flow is simulated using the open-source finite-volume-based solver OpenFOAM [83] adopting the RANS approach to model the turbulence. Accurate results are obtained for the target pressure gradient regardless of being constant or varying in the streamwise direction. Different numbers of design parameters 2, 4, and 8 are considered for which 25, 50, and 80 simulations are respectively needed to find the corresponding global optimum. The results of the optimum configuration for the constant-PG flows show an excellent agreement with the desired pressure-gradient distribution, and as consequence also with the correlations and experimental data reported in the literature for integral quantities of pressure-gradient boundary layers. In addition, promising results for the non-constant target PG distribution suggests that the BO-GPR approach can be applied to various practical applications including the design of inserts in wind tunnels. This is of special interest noting that the current approach for such designs is based on trial and errors, with limited accuracy and robustness.

Finally, we also apply the Bayesian optimization to an industrially-relevant case, in an effort to show that the approach can readily be used in an applied context. The goal is to optimize the configuration of the spoiler ice which is a simplified yet useful model for the actual ice profiles formed on the airfoils in cold conditions. The objective is to match, within a small margin of error, the resulting pressure distribution of an airfoil with modeled ice with that of the same airfoil but with actual ice profile. The RANS simulations using OpenFOAM for design parameters of dimension 5 and 7 are examined for which the maximum number of iterations in the optimization is found to be equal to 50 and 90, respectively. The encouraging results of the optimization prove the applicability of the BO-GPR framework to a relatively complex parameter optimization and also re-confirm the validity of the spoiler-ice models. It is also observed that keeping the heights in the spoiler-ice model fixed, as it is convenient in the literature, would lead to inaccurate flow simulations. Moreover, although the spoiler-ice concept has been mainly used in the literature for modeling the horn ice, see Refs. [51–53], the present study reveals that the model can also be applicable to the streamwise ice, conditioned on adopting the optimal parameters.

The present study can be extended in several ways. Our future attempts will include the use of scale-resolving approaches for simulation of the turbulent flows, instead of RANS which has been adopted here. In this way, even the effect of time-averaging on transient data can be considered. In addition, larger numbers of design parameters could be inquired, to identify the impact of dimensionality on the complete optimization problem. To make the BO more affordable, at least two strategies can be considered. As it is suggested for instance in Ref. [42], linear or non-linear mappings (feature mappings) can be introduced to map the high-dimensional space of the design parameters to a reduced-dimension space where the optimization of the acquisition function can be performed. As the second remedy, multifidelity modeling (MFM), see Refs. [56,49,60], can be combined with BO. In MFM, an accurate surrogate of the objective function can be constructed combining a large number of low-fidelity CFD simulations with only a few high-fidelity ones.

**Declaration of competing interest**

The authors declare that they have no known competing financial interests or personal relationships that could have appeared to influence the work reported in this paper.

**Acknowledgements**

YM acknowledges the Keio-KTH double degree program and the financial support from the NSK Scholarship Foundation. SR acknowledges the financial support from the Linné FLOW Centre at KTH and the EXCELLERAT project which has received funding from the European Union's Horizon 2020 research and innovation programme under grant agreement No. 823691. PS, NT and SR also acknowledge funding by the Knut and Alice Wallenberg Foundation via the KAW Academy Fellow programme No. 2018.0151. KF acknowledges the financial support from the Japan Society for the Promotion of Science (KAKENHI grant numbers: 18H03758 and 21H05007). The simulations in Section 6 were performed on the resources provided by the Swedish National Infrastructure for Computing (SNIC) at PDC (KTH Royal Institute of Technology), partially funded by the Swedish Research Council through grant agreement No. 2018-05973.

**Appendix A. Convergence of the BO-GPR**

According to the discussions in Section 2, the convergence of the BO can be traced by looking at the improvement of the surrogate of the objective in the parameter space and also checking the invariability of the best-value of the objective over sufficient number of samples. These two measures have been illustrated for the optimization problems in Sections 3 to 5. Figs. A.15, A.17, and A.19 show the improvement of the predicted mean and 95% confidence interval for the optimization of the toy problem (11), cavity flow, and turbulent boundary layer with pressure gradient. For the same set of problems, the variation of $r(\mathbf{q}^\dagger)$ is shown in Figs. A.16, A.18, and A.20. These plots show the improvement of the surrogate with increasing the number of the samples. This is also clear from the reduced uncertainty in the predictions near the optimum where sufficiently enough number of samples have been taken.

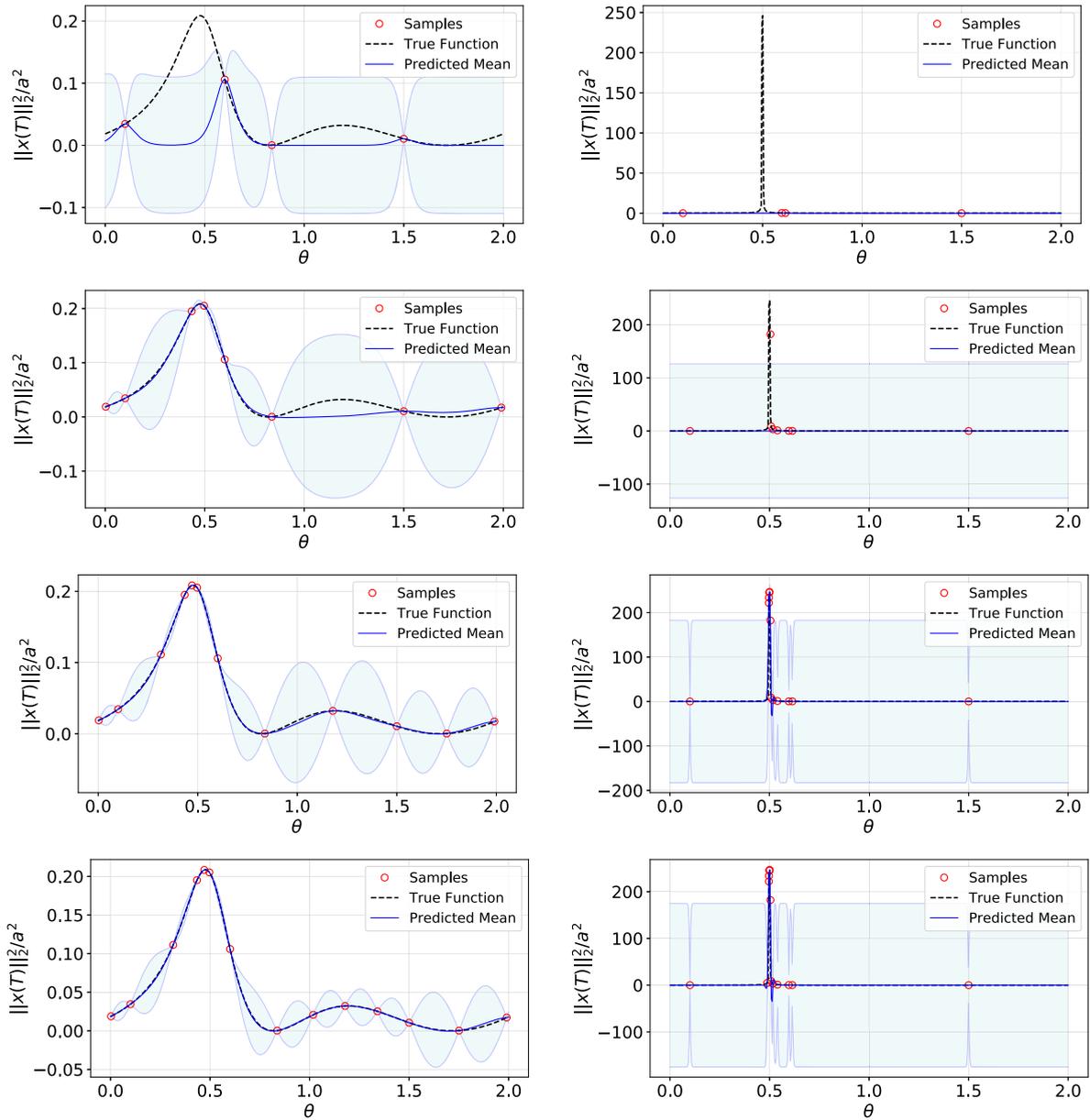

**Fig. A.15.** Impact of increasing the number of samples on the surrogate of the objective function in the optimization problem of Section 3. The plots are the same as those presented in Fig. 1 with (left) $a = 0.9$ and (right) $a = 1.0001$. The number of samples increases from 4 to 8, 12, and 14 from the top to the bottom row.

## Appendix B. Cost of the BO-GPR compared to CFD

At each iteration, the computational cost of the BO-GPR is mainly due to two main sources: optimization of the hyperparameters of the GPR and maximization of the acquisition function for finding next parameter sample. Clearly, as the number of design parameters and iterations increase, the cost of the BO-GPR grows. Based on our observations, the computational time for the BO-GPR is much smaller than a CFD run (order of $10^{-4}$ for the optimization of the spoiler-ice model in Section 6). In order to see if the cost ratio changes for larger number of parameters and higher number of iterations, minimization of the following algebraic equation can be considered,

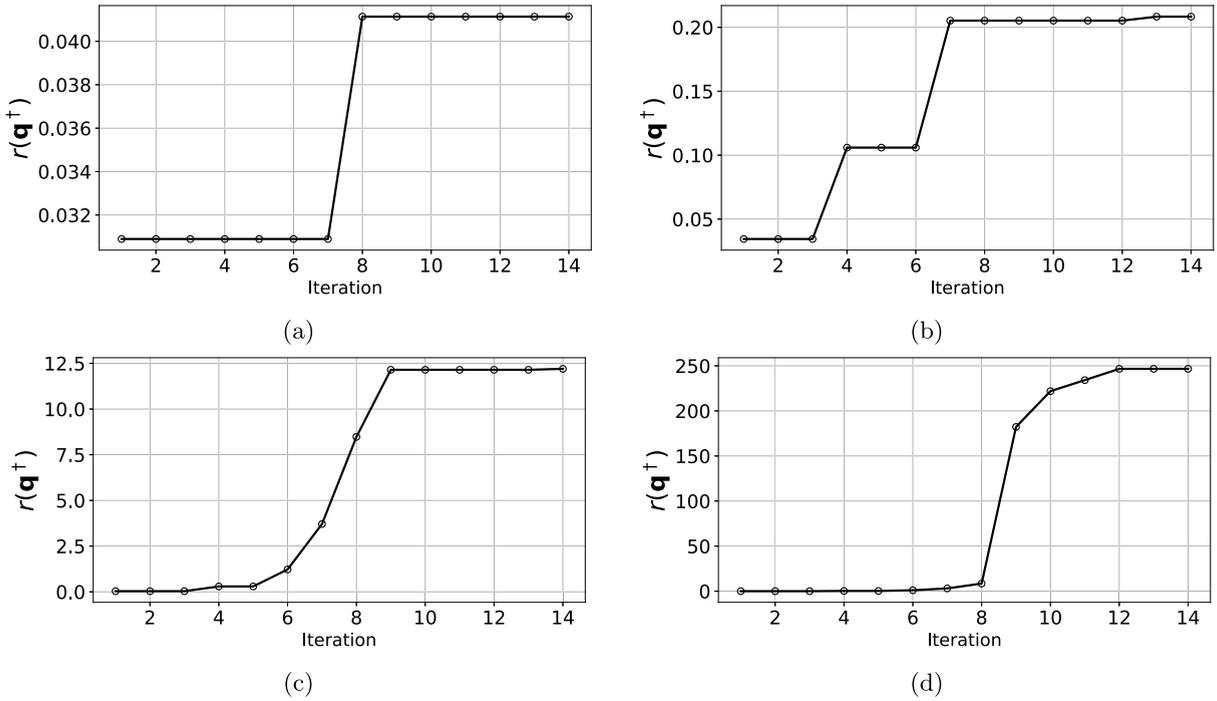

**Fig. A.16.** Convergence of the samples in the optimization problem of Section 3, with (a) $a = 10^{-4}$, (b) $a = 0.9$, (c) $a = 0.9999$, and (d) $a = 1.0001$. These plots are corresponding to the optimizations shown in Fig. 1.

$$r(\mathbf{q}) = \sum_{i=1}^{d} i \cdot q_i^i, \quad \text{where } q_i \in [-1, 1]. \tag{B.1}$$

The computational time for the BO-GPR is measured for three different numbers of parameters and up to 100 iterations. The results are reflected in Fig. B.21. Clearly, the computational time does not grow significantly with the number of parameters and the iteration. Although this observation can be dependent on the problem at hand, our experiments in the present study showed that the computational time of the BO-GPR is entirely dominated by the CFD simulations.

**Acronyms**

| | |
|---|---|
| **ACF** | Acquisition function |
| **APG** | Adverse pressure gradient |
| **BFGS** | Broyden-Fletcher-Goldfarb-Shanno |
| **BO** | Bayesian optimization |
| **CDF** | Cumulative distribution function |
| **CFD** | Computational fluid dynamics |
| **DNS** | Direct numerical simulation |
| **EI** | Expected improvement |
| **GEK** | Gradient-enhanced Kriging |
| **GLL** | Gauss-Lobatto-Legendre |
| **GPR** | Gaussian process regression |
| **LES** | Large eddy simulation |
| **LIC** | Line integral convolution |
| **MTL** | Minimum-turbulence-level |
| **PDF** | Probability density function |
| **PG** | Pressure gradient |
| **QoI** | Quantity of interest |
| **RANS** | Reynolds-averaged Navier–Stokes |
| **RSM** | Response-surface method |
| **SST** | Shear-stress transport |
| **TBL** | Turbulent boundary layer |
| **UQ** | Uncertainty quantification |
| **ZPG** | Zero pressure gradient |

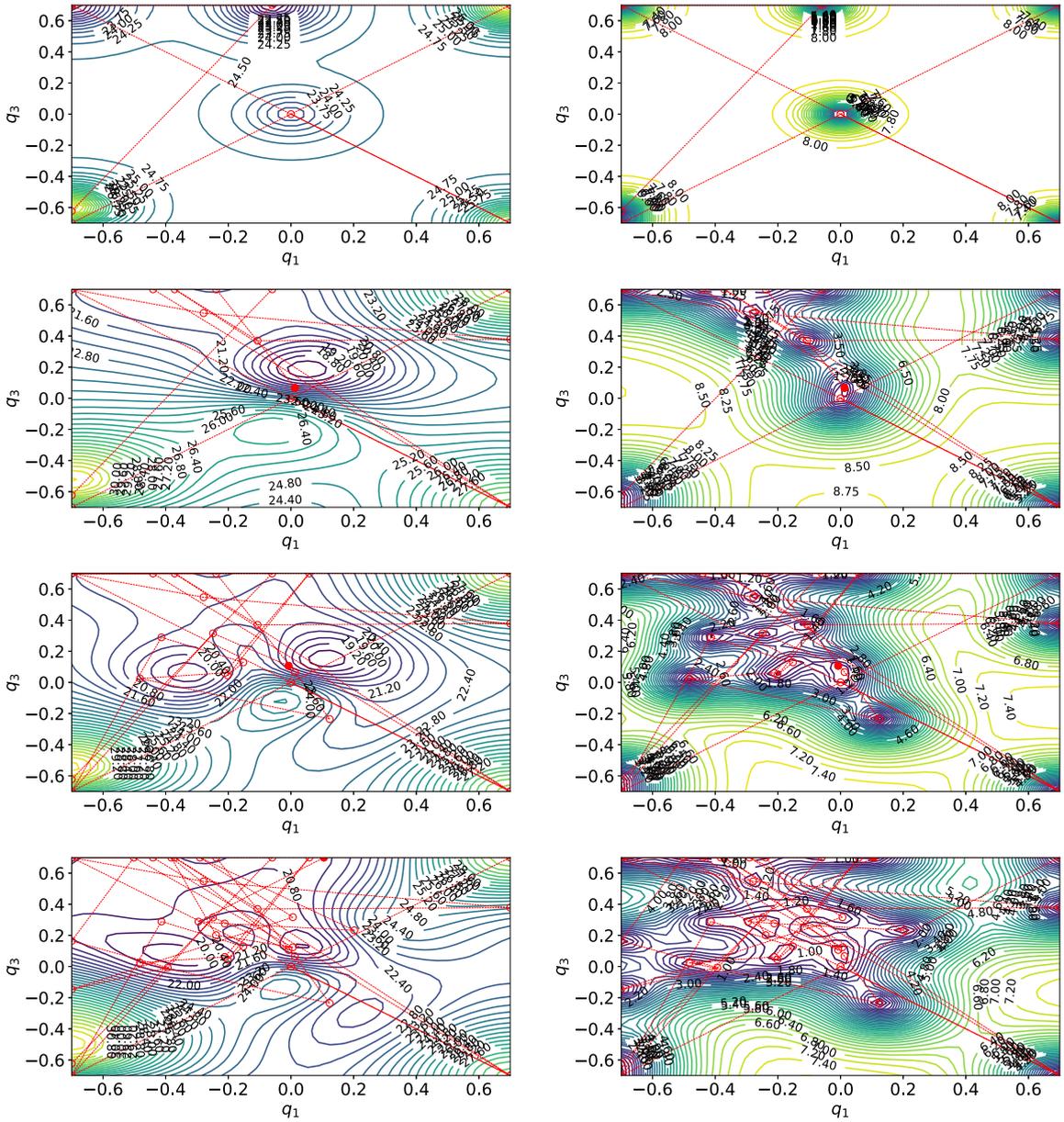

**Fig. A.17.** Impact of the number of samples on the (left) mean and (right) 95% confidence interval of the objective in the minimization problem of the cavity flow shown in Fig. 2. The surfaces are plotted in the space of $a_{1,L} - a_{1,R}$ parameters (labeled by $q_1$ and $q_3$, respectively). The number of samples increases from 10 to 20, 30, and 43 from the top to the bottom row.

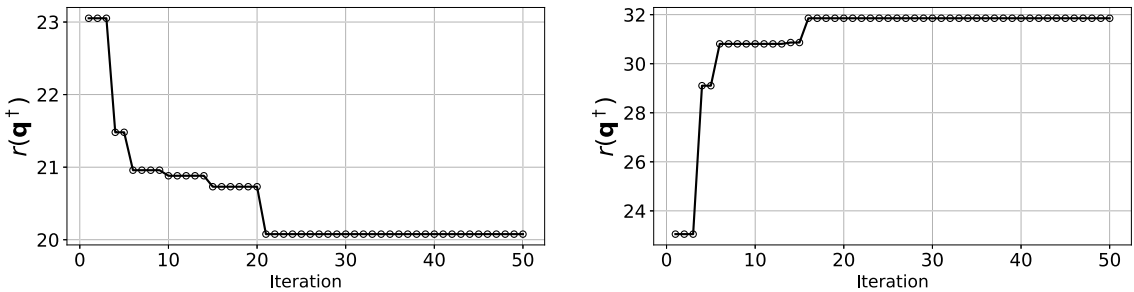

**Fig. A.18.** Convergence of the samples in the BO-GPR for (left) minimization and (right) maximization of the dissipation in the cavity flow at Re = 2000. These plots are associated with the optimization problems represented in Fig. 2.

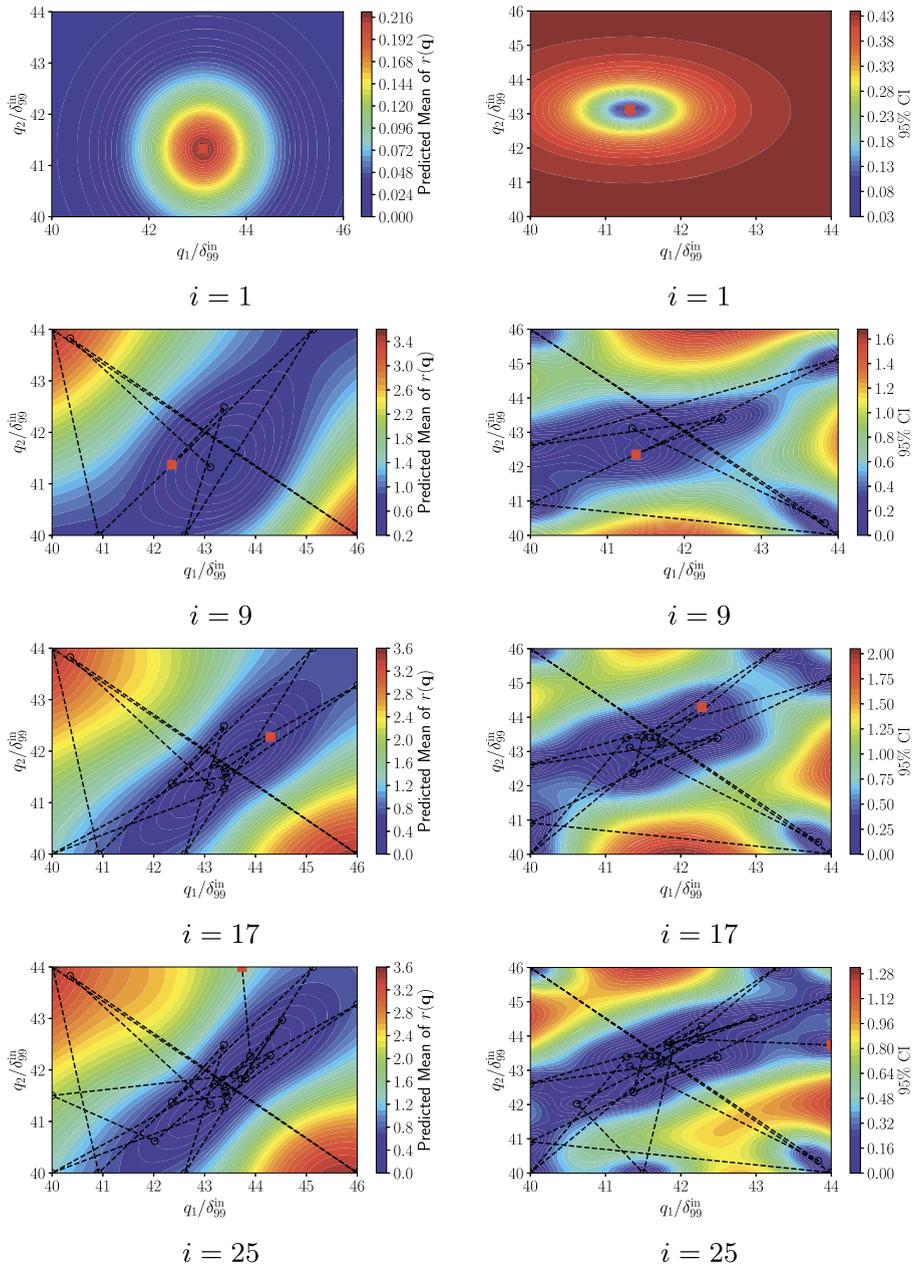

**Fig. A.19.** Impact of increasing the number of samples on the surrogate of the objective function for the Constant-0 case in Section 5. The plots are associated with those presented in Fig. 4.

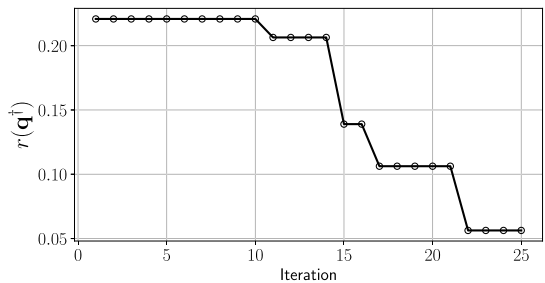

**Fig. A.20.** Convergence of the samples in the BO-GPR for the Constant-0 case in Section 5. The plot is corresponding to the optimizations shown in Fig. 4.

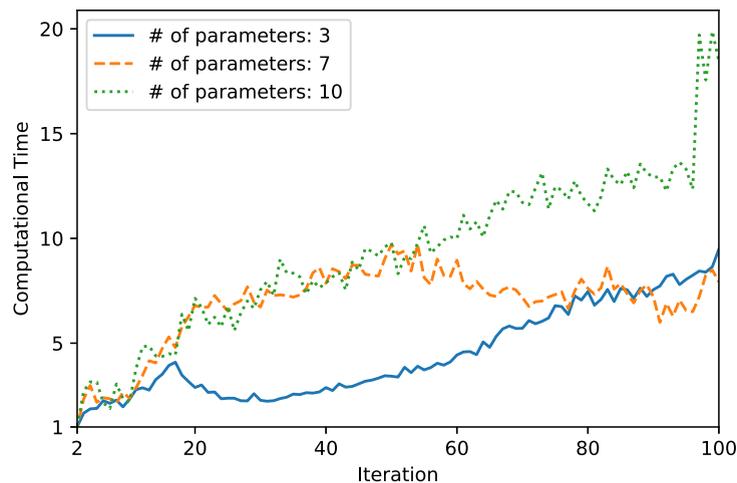

**Fig. B.21.** The computational time for a BO-GPR with different number of parameters and up to 100 iterations. The minimization of the algebraic equation (B.1) is considered. The computational time is averaged over 10 simulations for each number of parameters. Note that for each set of optimization, the computational time on the vertical axis is normalized with respect to the corresponding time needed for the second iteration, since the first sample is taken randomly. The time needed for the second iteration for 3, 7, and 10 parameters is 0.40, 0.46, and 0.48 seconds, respectively.